\documentclass[11pt,a4paper]{article}
\usepackage{jheppub,bm,color}

\usepackage{amsmath,amssymb,bm}

\newcommand{\na}{{\bm \nabla}}

\newcommand{\xv}{{\bm x}}
\newcommand{\yv}{{\bm y}}

\newcommand{\qv}{{\bm q}}

\newcommand{\rma}{{\rm a}}
\newcommand{\rmr}{{\rm r}}

\newcommand{\e}{{\rm e}}
\renewcommand{\i}{{\rm i}}

\def \pd {\partial}
\def \ep {\epsilon}

\def \le {\left}
\def \ri {\right}

\def \vx {\bm{x}}
\def \vy {\bm{y}}
\def \vk {\bm{k}}
\def \vq {\bm{q}}
\def \vnabla {\bm{\nabla}}

\def \G {\Gamma}
\def \eff {\rm eff}

\def \sG {{\cal G}}
\def \sA {{\mathcal A}}
\def \sL {{\cal L}}

\def \b {\beta}

\def \d {\delta}
\def \g {\gamma}

\def \o {\omega}

\def \l {\lambda}

\def \CC {{\cal C}}

\def \sC {{\cal C}}

\def \sP {{\cal P}}

\def \no {\nonumber}
\def \bes {\begin{subequations} }
\def \ees {\end{subequations}}

\def \<{\langle}
\def \>{\rangle}

\def \[{\left[}
\def \]{\right]}

\def \barW {\overline{W}}

\def \tI {t_{\rm I}}

\def \con {\rm c}
\def \equt {\rm equ.}

\def \tn {\tilde{n}}
\def \tpsi {\tilde{\psi}}
\def \txi {\tilde{\xi}}

\newcommand{\rmd}{{\rm{d}}}

\usepackage[normalem]{ulem}  % \sout{old text} 

\usepackage{tikz}
\usetikzlibrary{decorations.markings,decorations.pathmorphing,arrows.meta}
\tikzset{aux/.style={decorate,decoration={snake,segment length=1.5mm,
      amplitude=0.4mm}}}
\tikzset{left/.style={arrows={Stealth[scale length=0.5, scale width=1.5][sep=2pt]-}}}
\tikzset{right/.style={arrows={-[sep=-2pt]Stealth[scale length=0.5, scale
      width=1.5]}}}

\begin{document}
\title{
Off-equilibrium non-Gaussian fluctuations near the QCD critical point: an  effective field theory perspective
}
\author[a]{Noriyuki~Sogabe,}
\author[a]{and Yi Yin}

\affiliation[a]{Quark Matter Research Center, Institute of Modern Physics, Chinese Academy of Sciences, Lanzhou, Gansu, 073000, China }

\abstract{
The non-Gaussian fluctuations of baryon density are sensitive to the presence of the conjectured QCD critical point. 
Their observational consequences are crucial for the ongoing experimental search for this critical point through the beam energy scan program at Relativistic Heavy Ion Collider (RHIC). 
 In the expanding fireball created in a heavy-ion collision, critical fluctuations would inescapably fall out of equilibrium, and require a systematic description within a dynamical framework. 
 In this paper, we employ newly developed effective field theory (EFT) for fluctuating hydrodynamics to study the real-time critical non-Gaussian fluctuations of a conserved charge density. 
 In particular, we derive the evolution equations for multi-point correlators of density fluctuations and obtain the closed-form solutions with arbitrary initial conditions that can readily be implemented in realistic simulations for heavy-ion collisions. 
 We find that non-linear interactions among noise fields, which are missing in traditional stochastic hydrodynamics, could potentially contribute to the quartic (fourth order) fluctuations in scaling regime even at tree level in off-equilibrium situations.
}

\emailAdd{nori.sogabe@gmail.com}
\emailAdd{yiyin@impcas.ac.cn}

\maketitle

\section{Introduction}

The search for the conjectured QCD critical point has triggered much recent interest in both theory and experiment. 
This critical point is the endpoint of the first-order transition line, which separates the ordinary hadronic matter and quark-gluon plasma (QGP)~\cite{Stephanov:1998dy,Stephanov:1999zu}, and is considered as one of the landmarks on the phase diagram of strong interaction matter~\cite{Aprahamian:2015qub,Busza:2018rrf,Bzdak:2019pkr}. 
Many QCD-inspired models predict the existence of this critical point but disagree widely on its location (see ref~\cite{Fukushima:2010bq} for a reivew). 
Investigating the QCD critical point from the first principle lattice QCD calculation is currently formidable due to the sign problem at finite baryon density. 
The future development in quantum computing may help solve this sign problem someday (e.g.,~refs~\cite{Ikeda:2020agk,Honda:2021ovk}), but this is not today. 
On the observational frontier, looking for the QCD critical point is one of the main goals of the second phase of the beam energy scan (BES) program at Relativistic Heavy Ion Collider (RHIC) at Brookhaven National laboratory~\cite{Bzdak:2019pkr}. 
In addition, there are several other approved experiments anticipated in the coming years, such as the multipurpose detector (MPD) at the Nuclotron-based Ion Collider Facility (NICA) in Dubna and the compressed baryonic matter (CBM) experiment at the Facility for Antiproton and Ion Research (FAIR).% at GSI. 

The experimental exploration of the QCD critical point decisively relies on the observational consequences of baryon density fluctuations which are enhanced with the growth of the critical correlation length $\xi$. 
In equilibrium, their behaviors are fully determined by the equation of state (EoS) with a critical point.
However, those fluctuations would inescapably fall out of equilibrium in heavy-ion collisions because of the expansion of the fireball and the critical slowing down. There is mounting evidence indicating the off-equilibrium fluctuations differs from the equilibrium expectation not only quantitatively but also qualitatively ~\cite{Berdnikov:1999ph,Mukherjee:2015swa,Yin:2018ejt,Nahrgang:2018afz}.
This entails the development of dynamical models for density fluctuations;
see ref.~\cite{An:2021wof} which summarizes the current status of the dynamical modelling for BES physics.

The appropriate theoretical framework that serves this purpose is relativistic fluctuating hydrodynamics~\cite{Kapusta:2011gt}.
In the conventional approach, a la, Landau-Lifshitz, 
one adds white noise into hydrodynamic equations and fixes their magnitude by imposing fluctuation-dissipation theorem. 
Besides simulating stochastic hydrodynamics directly, one can alternatively derive the time evolution equation for hydrodynamic correlators~\cite{Akamatsu_2017,Stephanov:2017ghc,Martinez:2018wia,An:2019csj}. 
For example, such an equation for the two-point correlator of density fluctuations in systems relevant to the QCD critical point search has been investigated based on Hydro-Kinetic~\cite{Akamatsu:2018vjr} and Hydro+
formalism~\cite{Stephanov:2017ghc,Rajagopal:2019xwg,Du:2020bxp}.

The primary goal of the present work is to study the dynamics of non-Gaussian fluctuations of a generic conserved density near a critical point characterized by higher point correlators. 
Our study is motivated by the significant sensitivity of the non-Gaussian fluctuations to the growth of the critical correlation~\cite{Stephanov:2008qz,Asakawa:2009aj,Stephanov:2011pb}. 
In fact, 
the measurement of non-Gaussianity in the fluctuation of produced proton multiplicities as a function of beam energy is considered as one of the key observables for the criticality~\cite{STAR:2020tga,STAR:2021iop}. 
Recently, there has been significant progress in describing the real-time evolution of non-Gaussian fluctuations.
This includes the formulation of the evolution equation for higher point correlators of spatial-averaged order parameter field~\cite{Mukherjee:2015swa,Mukherjee:2016kyu} and diffusive mode~\cite{An:2020vri}, as well as the simulations of stochastic hydrodynamics~\cite{Nahrgang:2018afz}.

In contrast to existing works,
in this paper, we consider newly developed effective field theory (EFT) on the Schwinger-Keldysh (SK) contour for fluctuting fluid~\cite{Crossley:2015,Crossley:2017}, see also refs.~\cite{Kovtun:2014hpa,Haehl:2015foa,Jensen:2017kzi}. 
This EFT is constructed from action principle and symmetry.
Given the difference in this formalism from the other approaches used in the above-mentioned works, 
it would be useful to revisit the dynamics of the higher point hydrodynamic correlators using this EFT method.
Such a study will shed light on to what extent this new formalism would complement the existing ones. 
Particularly, 
the analysis so far has been limited to the Gaussian noise. 
While this is sufficient to generate non-Gaussian density fluctuations in equilibrium with an appropriate critical EoS~\cite{Mukherjee:2015swa,Nahrgang:2018afz,An:2020vri}, 
the effects of non-linear interactions among noises on off-equilibrium non-Gaussian fluctuations remain unclear to date.

We have applied the familiar field theory techniques for deriving the Schwinger-Dyson equation to obtain the tree-level evolution equations for multi-point fluctuations of a generic conserved density which is assumed to be the order parameter near a critical point, see eq.~\eqref{Wn-evo} together with eqs.~\eqref{S2}, \eqref{S3-B}, and \eqref{S4-B}. 
We also derive the closed-form solutions to those equations with arbitrary initial conditions, \eqref{Wn-sol}, which are new in literature. 
Those results are ready for implementation in realistic simulations for heavy-ion collisions. 
We have included non-linear coupling among noise fields discussed recently by Jain-Kovtun~\cite{Jain:2020fsm}, see eq.~\eqref{L}. 
Those interactions are parametrized by a set of "stochastic coefficients," which do not enter into the hydrodynamic constitutive relation and are absent in stochastic hydrodynamics.  
But as pointed out in ref.~\cite{Jain:2020fsm}, they do contribute to the hydrodynamic behavior of four-point retarded density correlators. 
For Gaussian and cubic fluctuations, we recover the previous results on evolution equations~\cite{An:2020vri} (see also ref.~\cite{Mukherjee:2015swa}) in hydrodynamic regime, but our expressions also apply to the scaling regime of a critical point.
Remarkably, we find that,
depending on the critical behavior of ``stochastic coefficients,'' the missing non-linear couplings among noise fields in the traditional fluctuating hydrodynamic do contribute to off-equilibrium quartic fluctuations.

This work is organized as follows. In section~\ref{sec:EFT}, 
we review the construction of EFT for a generic conserved density~\cite{Crossley:2015,Jain:2020fsm}. Then, we define the multi-point correlators for density fluctuations in section~\ref{sec:Wn-def} and derive their evolution equation following the textbook method of obtaining Schwinger-Dyson equation in section~\ref{sec:SD}. 
After introducing two small expansion parameters, we show evolution equations for Wigner transformed correlator at tree-level in section~\ref{sec:evo} and acquire the closed-form solutions to those evolution equations in section~\ref{sec:sol}. 
Section~\ref{sec:summary} is devoted to discussing the application of our result to the QCD critical point and related future directions.

\section{Formalism
\label{sec:formalism}
}

\subsection{EFT for a conserved density
\label{sec:EFT}
}

In this section, we review the ingredients of EFT for fluctuating hydrodynamics pertinent to the present study. 
The interested reader might refer to the lecture notes~\cite{Glorioso:2018wxw} for a pedagogical introduction.

Consider a many-body system described by a given microscopic theory.
One could imagine formally integrating out fast modes in the long-wavelength and low-frequency limit.
As a result, the system should be described by an effective action with a set of slow variables including conserved charge densities, i.e., hydrodynamic modes. 
This work will focus on EFT for a generic conserved charge density $n$ and assume $n$ becomes the order parameter near a critical point.

We can define this EFT action $I$ through the path integral representation of the generating functional $W$ on the Schwinger-Keldysh (SK) contour,
\begin{align}
\label{I-Z}
Z=\e^{{\rm i} W[A_\rmr,A_\rma]} = 
\int {\cal D} \psi_{\rm r}   {\cal D} \psi_{\rm a}
\e^{\i I[\psi_\rma,\psi_\rmr;A_{\rm a},A_{\rm r}] } \,,
\end{align}
where $I$ is a functional of the external gauge fields $A_{\rm r}$ and $A_{\rm a}$, and the dynamical fields $\psi_{\rm r}$ and $\psi_{\rm a}$. In SK formalism, 
the $\rm r$-variables are related to the physical observables (the charge density $n$ in the present case), while the $\rm a$-variables are the corresponding noise variables. One can also interpret $\psi_{\rmr}$ and $ \psi_{\rma}$ as the $\rm{U}(1)$ phase degrees of freedom on each fluid element (see refs.~\cite{Crossley:2015,Glorioso:2018wxw} for more details).

The EFT action $I$ and the associated Lagrangian density $\sL$ must satisfy various symmetry and consistency requirements, as we now explain: 
\begin{enumerate}
\item The action $I$ is invariant under the gauge transformation parametrized by an arbitrary ${\rm U}(1)$ phase  $\phi_{\rm r/a}$,
\begin{align}
\label{gauge}
A_{{\rm r/a},\alpha} \to A_{{\rm r/a},\alpha} - \pd_{\alpha}\phi_{\rm r/a}\, ,
\quad
\psi_{\rm r/a} \to \psi_{\rm r/a} +\phi_{\rm r/a} \, . 
\end{align}
Therefore, $I$ should be constructed from the gauge invariant combination: 
\begin{align}
\mathcal A_{{\rm r/a},\alpha} = A_{{\rm r/a},\alpha} + \partial_{\alpha}\psi_{\rm r/a}\, . 
\end{align}
The chemical potential is gauge invariant and can be identified with the temporal component of ${\cal A}_{\rmr,\alpha}$:
\begin{align}
\label{mu-def}
\mu= \sA_{r,0}\, .
\end{align}

\item 
The EFT has a local shift symmetry: 
\begin{align}
\label{shift-sym}
\psi_\rmr \rightarrow \psi_\rmr + \zeta(\xv) \, . 
\end{align}
Physically, this is because each fluid element is free to make independent ${\rm U}(1)$ phase rotations for time-independent $ \zeta(\vx)$. 
This shift symmetries will be absent when the global ${\rm U}(1)$ symmetry is spontaneously broken (see refs.~\cite{Dubovsky:2011sj,Crossley:2015} for further details).

\item
Fluctuation-dissipation relation is implemented through the invariance of the action under the so-called dynamical Kubo-Martin-Schwinger (KMS) transformation \cite{Glorioso:2017fpd,Glorioso:2016gsa}. In the ``classical'' limit that quantum fluctuations are small compared with the thermodynamic fluctuations, the dynamical KMS transformation is defined as
\begin{align}
\label{KMS}
{\cal A}_{\rmr,\alpha} & \rightarrow \Theta {\cal A}_{\rmr,\alpha}\,,
\quad
{\cal A}_{\rma,\alpha} \rightarrow \Theta {\cal A}_{\rma,\alpha}+ \i\beta \Theta \pd_{t}{\cal A}_{\rmr,\alpha}\, .
\end{align}
Here $\Theta={\cal P}{\cal T}$ represents a $Z_{2}$ discrete parity transformation and $\b=T^{-1}$ with  $T$ being the temperature. %\blue{Note that the dynamical KMS symmetry is broken by the terms specifying the initial configurations of the charge density.}

\item 
The unitarity of the underlying system requires that (suppressing the Lorentz indices)
\begin{align}
\label{unitarity}
I[\psi_\rmr,A_\rmr;\psi_\rma=0,A_\rma=0]=0\,, \qquad
I^{*}[\psi_\rmr,A_\rmr;\psi_\rma,A_\rma]=-I[\psi_\rmr,A_\rmr;-\psi_\rma,-A_\rma]\,. 
\end{align}

\end{enumerate}

Given the above conditions, 
$\sL$ can be constructed from the double expansion in the a-fields and gradient. 
Because of the requirement by the unitarity \eqref{unitarity}, 
$\sL$ has to comprise at least one power of a-field. 
Up to fourth order in a-fields and gradient,
the resulting EFT Lagragian density reads~\cite{Crossley:2015,Jain:2020fsm}
\begin{align}
\label{L-EFT}
\sL &= n {\mathcal A}_{\rm r,0} - \l  {\bm{\mathcal{A}}}_{\rm r} \cdot (\beta\partial_t {\bm{\mathcal{A}}}_\rmr - \i {\bm{\mathcal{A}}}_\rma) \no \\
%&+ \i \vartheta_{1}(\sA_{{\rm a},i}\sA_{{\rm a},j})\le[ (\beta\pd_{t} \sA_{r,i})(\beta\pd_{t} \sA_{r,j})-\delta_{ij}(\beta\pd_{t} {\bm\sA}_{\rm r})^{2}\ri]-\i\vartheta_{2}{\bm \sA}^{2}_{\rm a}(\beta\partial_t {\bm{\mathcal{A}}}_\rmr - \i {\bm{\mathcal{A}}}_\rma)^{2}\,,\notag\\
&\quad +\i \vartheta_{1} \left[
({\bm{\mathcal{A}}}_{\rm a}\cdot \beta \partial_t {\bm{\mathcal{A}}}_{\rm r})^2-{\bm{\mathcal{A}}}_{\rm a}^2 (\beta \partial_t {\bm{\mathcal{A}}}_{\rm r})^2 \right]-\i\vartheta_{2}{\bm \sA}^{2}_{\rm a}(\beta\partial_t {\bm{\mathcal{A}}}_\rmr - \i {\bm{\mathcal{A}}}_\rma)^{2}\,,
\end{align}
where $\l$ is the conductivity and $\vartheta_{1,2}$ are additional ``stochastic coefficients'' characterizing non-linear interactions among the density and noise fields.%
\footnote{The $\vartheta_{1,2}$ defined in the present paper is different from those in ref.~\cite{Jain:2020fsm} by a factor of $T^{2}$.}
As usual, the current can be read from the variation of the action with respect to the external gauge field, e.g.,
\begin{align}
\label{jr}
J^{\alpha}_{\rmr} = \frac{\delta I_{\rm eff}}{\delta A_{\rma,\alpha}}\longrightarrow
J^{0}_{\rm r}=n\,, 
\qquad
{\bm J}_{\rm r}= - \l \na \left( \frac{\mu}{T} \right)+ 2 \i  \l\na \psi_{\rm a}+\ldots\,,
\end{align}
where $\ldots$ denotes contributions from $\vartheta_{1,2}$. 
The first term in ${\bm J}_{\rm r}$ represents the familiar Fick's law, while the second term accounts for the effect of Gaussian noise.
The usual diffusive equation can be obtained by varying $\sL$ with respect to $\psi_{\rm a}$ and then setting all the a-fields to be zero.

In the absence of $\vartheta_{1,2}$, 
the EFT action~\eqref{L} is quadratic in a-field and can be matched with the traditional stochastic hydrodynamics for diffusive mode with white Gaussian noise~\cite{Glorioso:2018wxw}. 
The $\vartheta_{1}$-contribution to the Lagrangian density \eqref{L} is quadratic in noise field while $\vartheta_{2}$-contribution also contains cubic and quadratic interactions among noise fields.
Both contributions do not enter into the standard stochastic hydrodynamics. 
As a prior, the cubic and quartic interactions among noise fields might be parametrized by a handful of ``stochastic coefficients'' even at fixed order in gradient. 
The dynamical KMS symmetry fixes the form of those terms, as shown in ref.~\cite{Jain:2020fsm}. 
This demonstrates that one can employ the EFT approach to investigate non-linear noise-noise and noise-density interactions, systematically.

In what follows, we shall consider the EFT Lagrangian density~\eqref{L-EFT} at vanishing external field as our starting point to analyze the fluctuation dynamics of density $n$: 
\begin{align}
\label{L}
\mathcal{L} &= n \partial_t \psi - \lambda \na \psi \cdot (\beta\na \mu - \i \na \psi)
 \no \\ 
%&+\i \vartheta_{1}(\pd_{i}\psi)(\pd_{j}\psi)\le[(\beta\pd_{i}\mu)(\beta\pd_{i}\mu)-\delta^{ij}(\beta{\vnabla\mu})^{2}\ri]-\i\vartheta_{2}(\vnabla \psi)^{2}\,(\beta\vnabla\mu - \i\vnabla\psi)^{2}\,,
&\quad +\i \vartheta_{1}
\left[ (\na\psi \cdot \beta \na \mu)^2 - (\na\psi)^2 (\beta \na\mu)^2 \right] -\i\vartheta_{2}(\vnabla \psi)^{2}(\beta\vnabla\mu - \i\vnabla\psi)^{2}\,,
\end{align}
where we have used eq.~\eqref{mu-def}.
Here and hereafter, we drop the subscript for $\psi \equiv \psi_{\rm a}$.
The chemical potential $\mu$ and $n$ will be related by equation of state (EoS) which encodes the information on the critical point.

As mentioned earlier, our study will focus on the non-Gaussian fluctuations near a critical point. 
$\vartheta_{1,2}$ terms are at least fourth-order in gradient and can be safely dropped in the hydrodynamic regime $q\xi \ll 1$, where $q$ is the typical momentum of fluctuating density mode and $\xi$ is the critical correlation length. However, they could be potentially important for off-equilibrium non-Gaussian fluctuations in the scaling (critical) regime $q \xi \geq 1$, (c.f.,~eq.~\eqref{L4}).  
As we shall see below, 
those $\vartheta_{1,2}$-contributions indeed enter into the evolution equation for quartic density fluctuations even at tree-level.

\subsection{
Multi-point correlators
\label{sec:Wn-def}
}

We are interested in the density fluctuation around a given hydrodynamic background
\begin{align}
\label{background}
n(x)= (n(x))_{0}\,,
\qquad 
\psi=0\,,
\end{align} 
which corresponds to the saddle point of the action~\eqref{L}. 
Here, we use $x=(t,\vx)$ to denote space time coordinates and adopt the subscript $0$ for the background value. 
The background charge density satisfies the diffusive equation,
\begin{align}
\label{n-background}
\pd_{t} (n(x))_{0}= (F)_{0}(x)\,, 
\qquad
F(x)= - \vnabla\cdot {\bm J}_{\rm r}\,,
\end{align}
% where $F$ is given by the divergence of $r$-current~\eqref{jr}
% \begin{align}
% \label{F}
% F(x)= - \vnabla\cdot {\bm J}_{\rm r}
% %= \vnabla\cdot(\l\,\beta \vnabla\mu)-2i \vnabla\cdot(\l\vnabla\psi)+
% %\textrm{$\vartheta_{1,2}$ contributions}\, . 
% \end{align}

To describe the correlation among the fluctuations around the background profile,
\begin{align}
\delta n(x)=n(x)-(n(x))_{0}\,,
\end{align}
we define multi-point equal-time correlators, 
\begin{align}
\label{W-def}
W_{n}(t;\vx_{1},\ldots,\vx_{n})\equiv \left<  \prod^{n}_{i=1}\, \delta n(t,\vx_{i}) \right>_{\con}\,,
\end{align}
where the subscript ``\con'' denotes the connected contribution and the average is given by
\begin{align}
\langle \ldots \rangle =\frac{1}{Z} \int {\cal D} n {\cal D}\psi e^{\i I} (\ldots)\,.
\end{align}

\subsubsection{Equilibrium
\label{sec:equilibrium}
}

Before studying the real-time evolution of multi-point correlators~(\ref{W-def}), we here first review their properties in equilibrium. 
In the present subsection, we shall assume the hydrodynamic background is static and homogeneous.

We begin by considering the free-energy functional of the form,
\begin{align}
\label{E-a}
\beta {\cal E}[\delta n] = \int_\xv \left[ 
\frac{1}{2 }a (\delta n)^2+ \frac{1}{2 }K(\na \delta n)^2 
+\frac{1}{3!}a'  (\delta n)^{3}
+\frac{1}{4!}a'' (\delta n)^{4} \right] \,,
\end{align}
where $\int_\xv \equiv \int \rmd \xv$, $a=\beta\, \mu'$ with prime denoting the derivative with respect to $n$ with $T$ fixed, and $K$ corresponds to the surface tension. 
Then, the multi-point equal-time correlator in equilibrium (denoted by bar), $\barW_{n}$, is given by (c.f.~eq.~\eqref{W-def}),
\begin{align}
\label{W-equi}
\barW_{n}(\vx_{1},\ldots,\vx_{n}) = \left<  \prod^{n}_{i=1} \delta n(t,\vx_{i}) \right>_{\con,{\rm eq}}\,, 
\end{align}
where the equilibrium expectation can be evaluated as 
\begin{align}
\langle\ldots\rangle_{\rm eq} = \frac{1}{Z_{{\rm eq}}} \int {\cal D} \delta n e^{-\beta {\cal E}[\delta n]} (\ldots)\,, \qquad Z_{{\rm eq}}\equiv \int {\cal D} \delta n e^{-\beta {\cal E}[\delta n]}\,. 
\end{align}
Below, we shall consider $\barW_{n}$s in Fourier space,
\begin{align}
\label{Weq-Four}
\barW_{n}(\vq_{1},...,\vq_{n}) & = \int_{\yv_1} \cdots \int_{\yv_n} 
\barW_{n}(\vx_{1},\ldots,\vx_{n})
\delta\left(\frac{\vx_1+\cdots+\vx_n}{n} \right) \prod^{n}_{j=1}\e^{-\i \vq_{j}\cdot{\vx_{j}}} \, . 
%%%
\end{align}

For small magnitude of deviations away from the equilibrium,
we can evaluate eq.~\eqref{W-equi} by only accounting for ``tree-level'' contribution. 
More explicitly, 
for the equilibrium correlator $\barW_{2}$, 
we only keep Gaussian part of eq.~\eqref{E-a} and find that $\barW_{2}$ takes the familiar Ornstein-Zernike form,
\begin{align}
\label{W2-eqA}
    \barW_{2}(\vq) = \frac{1}{a f(\qv)}\, : \qquad 
    \raisebox{-12pt}{\tikz[thick]{
    \draw [dashed] (-14mm,0)--(-7mm,0);
    \draw [dashed](-7mm,0)--(0,0);
    \node at (-7mm,-3mm) {$\qv$};
    }}\,,
\end{align}
where
\begin{align}
\label{f-def}
     f(\qv)= 1+c \qv^{2}\, ,
     \qquad
      c= \frac{K}{a}= \xi^{2}\,, 
\end{align}
and the dashed line on (\ref{W2-eqA}) is the diagram corresponding to $\barW_{2}$. In this work, we shall replace the critical exponents $\eta\approx 0.04$ by $0$ for simplicity.

Turning to $\barW_{3}$, 
it receives the following contribution,
\begin{align}
\label{W3eq-in-diam}
    \barW_{3}(\vq_{1},\ldots,\vq_{3})&=\qquad
      \raisebox{-29pt}{\tikz[thick]{
    \draw [dashed] (-10mm,0)--(0,0);
    \node at (-5mm,-3mm) {$\vq_1$};
    \draw [dashed] (0mm,0mm)--(10mm,10mm);
    \node at (8mm,4mm) {$\vq_2$};
    \draw [dashed] (0mm,0mm)--(10mm,-10mm);
    \node at (8mm,-4mm) {$\vq_3$};
    \draw[fill] (0,0) circle (.6ex);
    %\node at (0,-8mm) {$(-a')$};
  }}\qquad
   = \qquad
    - a'\prod^{3}_{i=1}\barW_{2}(\vq_{i})\,, 
\end{align}
where the cubic vertex obtained from eq.~\eqref{E-a} is proportional to $-a'$. 
Meanwhile, 
$\barW_{4}$ is given by the contribution from the sum of the two diagrams:
\begin{align}
\label{W4-equil}
 %\begin{tikzpicture}
  \barW_{4}(\vq_{1},\ldots,\vq_{4})&= 
  \qquad
   %
    % First diagram
    %
  \raisebox{-26pt}{
  \tikz[thick]
  {
      \draw [dashed] (-10mm,10mm)--(0mm,0mm);
      \node at (-8mm,4mm) {$\vq_1$};
      \draw [dashed] (-10mm,-10mm)--(0,0);
      \node at (-8mm,-4mm) {$\vq_2$};
      \draw [dashed] (10mm,10mm)--(0,0);
      \node at (8mm,4mm) {$\vq_3$};
      \draw [dashed] (10mm,-10mm)--(0,0);
      \node at (8mm,-4mm) {$\vq_4$};
      \draw[fill] (0,0) circle (.6ex);
      %\node at (0,-4mm) {$a''$};
    }
    }
    \qquad
    +
    \qquad
    %
    % Second diagram
    %
      \raisebox{-29pt}{\tikz[thick]{
      %\draw[aux] (-10mm,10mm)--(-5mm,5mm)
      %\draw[aux] (-5mm,5mm)--(-10mm,0)
    \draw [dashed] (-10mm,10mm)--(0mm,0mm);
           \node at (-8mm,4mm) {$\vq_1$};
      \draw [dashed] (-10mm,-10mm)--(0,0);
            \node at (-8mm,-4mm) {$\vq_2$};
      \draw [dashed] (0,0)--(10mm,0);
      \draw [dashed] (10mm,0)--(20mm,10mm);
                  \node at (18mm,4mm) {$\vq_3$};
      \draw [dashed] (10mm,0)--(20mm,-10mm);
         \node at (18mm,-4mm) {$\vq_4$};
         \draw[fill] (0,0) circle (.6ex);
         \draw[fill] (10mm,0) circle (.6ex);
%    \draw [aux] (-5mm,0)--(0,0);
%    \node at (-5mm,-3mm) {$p_1$};
%    % 
%    \draw [right] (0mm,0)--(5mm,5mm);
%    \draw [aux] (5mm,5mm)--(10mm,10mm);
%    \node at (8mm,4mm) {$p_2$};
%    % 
%    \draw [right] (0mm,0)--(5mm,-5mm);
%    \draw [aux] (5mm,-5mm)--(10mm,-10mm);
%    \node at (8mm,-4mm) {$p_3$};
  }}
%\end{tikzpicture}
\nonumber
\\
 &= \le[
    -a''+ 
    (a')^{2}\le(
    \barW_{2}(\vq_{12})+\barW_{2}(\vq_{13})+\barW_{2}(\vq_{14})
    \ri)
    \ri] \prod^{4}_{i=1}\barW_{2}(\vq_{i})\,,
\end{align}
where the quartic vertex is proportional to $-a''$ and we have introduced the notation,
\begin{align}
\label{qij}
\vq_{ij}=\vq_{i}+\vq_{j}\,.
\end{align}

To this point, we have assumed that $a'$ and $a''$ have no dependence on the momenta $\vq_{i}$. 
Strictly speaking, for a system near the critical point, this is only true for $ |\qv_{i}|\xi \leq 1$. 
In the scaling regime $|\qv_{i}|\xi \geq 1$, $a'$ and $a''$ can strongly depend on $\vq_{i}$. 
We defer the investigation of this subtle point to future studies.

\subsection{Schwinger-Dyson equations
\label{sec:SD}
}
In this section, 
we shall follow the textbook derivation of Schwinger-Dyson equations from the path integral to obtain the analogous equations for $W_{n}$.

We begin by considering the one-point function: 
\begin{align}
\label{n1}
\langle \delta n(x)\rangle =\frac{1}{Z} \int {\cal D} n {\cal D}\psi e^{\i I} \delta n(x)\,,
\end{align}
and replace $\psi(x)\to \psi(x)+\epsilon(x)$ with $\epsilon(x)$ being an arbitrary local function with a small magnitude. 
Since this is just a field redefinition and the path integral integrates over all configurations, the L.H.S of eq.~\eqref{n1} should be the same. 
Therefore, the change of the R.H.S. of eq.~\eqref{n1} must add to zero. 
At first order in $\epsilon(x)$, we arrive at the condition
\begin{align}
\left< \frac{\d I}{\d \psi(x_{1})} \delta n(x) \right> =0\,, 
\end{align}
or equivalently
\begin{align}
\left< \pd_{t_{1}} \delta n(x_{1})\delta n(x) \right>= \left< F(x_{1}) \delta n(x) \right>\, . 
\end{align}
By substituting eq.~\eqref{n-background} into the above equation, 
we have
\begin{align}
\langle  \pd_{t_{1}}n(x_{1}) \delta n(x)\rangle=  \langle  \delta F(x_{1})\,\delta n(x) \rangle\,,
\end{align}
where $\delta F=F- (F)_{0}$.
Similar steps give
\begin{align}
\label{dnn}
\left< \pd_{t_{1}}n(x_{1}) \prod^{n}_{j=2}\delta n(x_{j}) \right> = \left< \delta F(x_{1}) \prod^{n}_{j=2}\delta n(x_{j}) \right>\,. 
\end{align}

Now, we are ready to obtain the evolution equation for $W_{n}$.
For this purpose, we consider
\begin{align}
\label{dG}
\sum^{n}_{i=1} \pd_{t_{i}}\left< \prod^{n}_{i=1}
\delta n(x_{i}) \right>_{\con}\,,
%    =\frac{1}{n!}\sum^{n}_{i=1}\le[\langle F(x_{i})\, \prod_{j\neq i}\, \delta n(x_{j})\rangle_{c}\ri]_{\sP_{\{\vx_{n}\}}}\,
\end{align}
where we can view $\langle \delta n(x_{i})\rangle_{\con}$ as a function of
\begin{align}
t=\le(t_{1}+\ldots+t_{n}\ri)/n \,,\qquad \Delta t_{1}=t_{1}-t \,,\quad ...\,,\quad  \Delta t_{n-1}=t_{n-1}-t \,.
\end{align}
%$t=\le(t_{1}+\ldots+t_{n}\ri)/n,\Delta t_{1}=t_{1}-t,\ldots, \Delta t_{n-1}=t_{n-1}-t$. 
Noting $ \pd_t=\sum^{n}_{i}\pd_{t_{i}}$, 
we conclude that taking the equal-time limit $t_{1},\ldots,t_{n}\to t$ on eq.~\eqref{dG} gives $\pd_{t}W_{n}(t;\vx_{1},\ldots,\vx_{n})$.
On the other hand, 
eq.~\eqref{dG} can be evaluated using eq.~\eqref{dnn}, 
leading to the evolution equation:
\begin{align}
\label{SE-eq}
    \pd_{t}W_{n}(t;\vx_{1},\ldots,\vx_{n})%\notag\\
    &=\lim_{\equt}\sum^{n}_{i=1}
    \left< F(x_{i}) \prod_{j\neq i} \delta n(x_{j}) \right>_{\con}
    \nonumber\\
    &=\frac{1}{2}\lim_{t\to t'} 
     \le[
     \left< F(t,\vx_{1})\, \prod^{n}_{j=2}\, \delta n(t',\vx_{j}) \right>_{\con}
     +\left<  F(t',\vx_{1})\, \prod^{n}_{j=2}\, \delta n(t,\vx_{j})\right>_{\con}
     \ri]_{\CC_{\vx_{1},\ldots,\vx_{n}}}\,,
\end{align}
where $\sC_{\vx_{1},\ldots,\vx_{n}}$ denotes the n-cyclic permutation of $(\vx_{1},\ldots,\vx_{n})$.
For example, for a generic function $w(\vx_{1},\vx_{2},\vx_{3})$, the operation gives
\begin{align}
\label{cyc_def}
\le[w(\vx_{1},\vx_{2},\vx_{3})\ri]_{\sC_{\vx_{1},\ldots,\vx_{3}}}
=w(\vx_{1},\vx_{2},\vx_{3})+w(\vx_{3},\vx_{1},\vx_{2})+w(\vx_{2},\vx_{3},\vx_{1})\,.
\end{align}
On the last expression of eq.~\eqref{SE-eq}, 
we have specified the prescription for taking the equal-time limit such that the results will not depend on whether $t-t'\to 0^{+}$ or $t-t'\to 0^{-}$.

Before closing this section, we remark that the method presented in this section can equally apply to derive evolution equation for un-equal time correlators, see appendix.~\ref{sec:G-t} for illustrative examples.

\section{Evolution equations
\label{sec:evo}
}
\subsection{Perturbative scheme}
At this juncture, the Schwinger-Dyson equation~\eqref{SE-eq} is exact and non-perturbative. 
To make progress, 
we consider an perturbative evaluation of eq.~\eqref{SE-eq}. 

First, we expand the Lagrangian density~\eqref{L} in fluctuating fields $\delta n$ and $\psi$:
\begin{align}
  \sL= (\sL)_{1}+(\sL)_{2}+(\sL)_{3}+\ldots\, ,   
\end{align}
where the subscript of $(\ldots)$ counts the power of fluctuating fields.
Note $(\sL)_{1}$ is simply a total derivative for expansion around the saddle point of the action. 
Explicitly, we have from eq.~\eqref{L},
\begin{align}
\label{L2}
(\sL)_{2} &= - \psi \pd_{t}\d n +\psi \vnabla \cdot \le[ \g \vnabla\le(1- c \na^{2}\ri)\d n - \i  \l \vnabla\psi\ri]\,,\\
\label{L3}
(\sL)_{3}&= 
\psi\vnabla\cdot\le[
\frac{1}{2}\g' \vnabla (\d n)^{2} - ac \l' \delta n \na (\na^{2}\delta n)-\i \l' \d n \na \psi \ri] \,, \\
\label{L4}    
(\sL)_{4}&=\psi\vnabla\cdot \le[ \frac{1}{6}\g''\vnabla(\d n)^{3} -\frac{1}{2} ac \l'' (\delta n)^{2}\na (\na^2\delta n) - \frac{i}{2}\l'' (\d n)^{2} \vnabla\psi \ri] \nonumber\\
%&\quad +\i \vartheta_{1}(\pd_{i}\psi)(\pd_{j}\psi)a^{2} \le[(\pd_{i}(1-c\na^2)\delta n)(\pd_{j}(1-c\na^2)\delta n)-\delta^{ij} \le(\vnabla(1-c\na^2)\delta n\ri)^{2} \ri]\notag\\
&\quad +\i \vartheta_{1}a^{2} \le[(\na\psi \cdot \na(1-c\na^2)\delta n)^2 - (\na\psi)^2 \le(\vnabla(1-c\na^2)\delta n\ri)^{2} \ri]\notag\\
&\quad -\i\vartheta_{2}(\vnabla \psi)^{2}\,\le(a\vnabla(1-c\na^{2}\ri)\delta n
-\i \vnabla\psi)^{2}\, ,
\end{align}
where we have used eq.~\eqref{E-a} to give the relation between $\d\mu$ and $\d n$,
\begin{align}
\label{dh-a}
    \beta \delta \mu=\beta\frac{\delta {\cal E}}{\delta n}=a\, \le( 1- c\na^{2}\ri)\, \delta n + \frac{1}{2!}\,a' (\delta n)^{2}+\frac{1}{3!}a''(\delta n)^{3}\, .
\end{align}
In the above expressions, 
the diffusive constant is given by
\begin{align}
\label{diff}
    \g = a\, \l\, . 
\end{align}

The interacting Lagrangian density $\mathcal{L}_{\rm int} = (\sL)_{3}+(\sL)_{4}+\cdots $ describes non-linear interaction among $\delta n$ and noise field $\psi$. 
They receive contributions depending on the variations of inverse susceptibility $a$ and conductivity $\l$ with respect to thermodynamic quantities. 
The latter contribution, e.g., $\l',\l''$, is sometimes referred to as the multiplicative noise effects because $\l$ controls the magnitude of Gaussian noise in conventional stochastic hydrodynamics. %Its dependence on thermodynamics variables such as $n$ gives rise to a non-linear coupling between densities and noise fields, say $\propto \psi \psi \delta n$. 
Although those multiplicative noise contributions can involve two powers of fluctuating {\it density} fields in $\sL_{{\rm int}}$, to say $\propto \psi \psi \delta n \delta n$, they only involve two powers of {\it a-field} and do not capture the non-linear interaction among the noise fields. 
In contrast, $\vartheta_{2}$-contribution involves cubic and quartic power in noise-field, such as $\propto \psi \psi \psi \delta n$ and  $\propto \psi \psi \psi \psi $, so it represents a different class of noise-density interactions from multiplicative noise.

In hydrodynamic regime $q\xi \ll 1$, both surface tension (terms proportional to $c$) and $\vartheta_{1,2}$ contribution are unimportant as they involve four powers of gradient. 
However, in the scaling regime $q\xi \geq 1$, one can not even obtain the correct equilibrium expectation for $\barW_{2}$ without including $c$-contributions since $c\sim \xi^{2}$ and hence is not suppressed by $q\xi$~(see eq.~\eqref{W2-eqA}). 
Similarly, we expect the significance of $\vartheta_{1,2}$ contribution if they exhibit non-trivial $\xi$-dependence near the critical point. 
As shown in ref.~\cite{Jain:2020fsm}, 
$\vartheta_{1,2}$ can be extracted from four-point retarded correlator for any given microscopic theory. 
Currently, to the best of our knowledge, the critical behavior of ``stochastic coefficients'' $\vartheta_{1,2}$ remains elusive but deserves investigation in the future.

Next, we set up a power counting scheme. Following the standard procedure in EFT, we introduce the rescaled fields:
\begin{align}
\label{rescale}
    \delta \tn = g\delta n\, , 
    \qquad
    \tpsi = g^{-1}\psi\, , 
\end{align}
where $g^{-2}=a^{-1}$ is the equilibrium Gaussian fluctuation of $\delta n$ per unit volume (see eq.~\eqref{W2-eqA}). 
So, the Gaussian fluctuation of the rescaled variable $\delta \tn$ will be the order of the unity (c.f.~eq.~\eqref{W2-eqA}), 
which is one primary motivation for the definition~\eqref{rescale}. 
Even away from the critical point, $g$ would become small when the degree of freedom per phase space volume, i.e., entropy density, is large. 
For example, for the QGP, $g^{2}$ will be suppressed if the number of color charge $N_{c}$ and quark flavor $N_{f}$ become large. 
Near the critical point, 
$g^{2}$ is suppressed as $\xi^{-2}$ due to the growth of the correlation length $\xi$.
To make the dependence of $g$ on entropy density and on correlation length more transparent, 
we introduce the parametrization:
\begin{align}
\label{g0}
    g = g_{0} \txi^{-1}\,, 
    \qquad
    \txi \equiv \frac{\xi}{\xi_{0}}\,,
\end{align}
where $\xi_0$ is the typical microscopic correlation length.  
By design, we require $g_0$ to be small for systems with large entropy density but is insenstive to the correlation length.

By expressing $(\sL)_2$ in terms of the rescaled field~\eqref{rescale}, it is easy to check that $(\sL)_2\sim {\cal O}(g^{0})$ (note $\l\sim g^{-2}$). 
Turning to $(\sL)_{n>2}$, 
we first note that for a generic thermodynamic function $A$, 
\begin{align}
    \frac{A'}{A}= \frac{1}{A}\frac{\partial A}{\partial \mu}\left( \frac{\pd n}{\pd \mu}\right)^{-1}\sim g^2 \frac{1}{A}\frac{\partial A}{\partial \mu}\, .
\end{align}
Since $A^{-1}\partial A/\partial \mu $ should not depend on the density degrees of freedom, it can be counted as $O(1)$ in a small $g_0$ limit. 
On the other hand, if $A$ scales with correlation length near the critical point, so does its derivative with respect to $\mu$. 
For definiteness, 
we shall assume the critical point under consideration is in the 3d Ising model universality class so that $T$ and $\mu$ can be mapped to $r$ and $h$, where $r$ is the reduced Ising temperature and $h$ is the Ising magnetic field. 
In general, 
$\partial /\partial \mu$ can be expressed as a linear combination of $\pd_{h}$ and $\pd_{\rm r}$. Given that $\pd_{h}$ gives stronger dependence on the correlation length, we then have
\begin{align}
    \frac{1}{A}\frac{\partial A}{\partial \mu}\sim \frac{\pd_{h}A}{A} \sim \txi^{\frac{2\delta_c}{\delta_{c}-1}}\, , 
\end{align}
where $\delta_c\approx 5$ is the well-known Ising critical exponents. 
This leads to the relation
\begin{align}
     \frac{A'}{A} \delta n \sim  g \txi^{\frac{2\d_{c}}{\d_{c}-1}}\delta \tn =g_{0}\txi^{\frac{\d_{c}+1}{\d_{c}-1}} \delta \tn
    = g_{\eff}\, , 
\end{align}
where we have defined
\begin{align}
    g_{\eff}\equiv g_{0}(\txi)^{\frac{\d_{c}+1}{\d_{c}-1}}
    = g_{0}(\txi)^{\frac{3}{2}}\, . 
\end{align}
This implies that $\sL_{3}\sim (\pd \sL_{2}/\pd n)\delta n$ should be counted as $g_{\eff}$.
More generally, one can show by following the analogous steps that
\begin{align}
\label{sL-counting}
    \sL_{n}\sim g^{n-2}_{\eff}\, , 
\end{align}
meaning $g_{\eff}$ can be employed as an effective coupling constant organizing the loop corrections to the tree-level results. 
It also follows from eq.~\eqref{sL-counting} that
\begin{align}
    \left< \prod^{n}_{i=1}\delta \tn \right>_{\con}\sim g^{n-2}_{\eff}\, , 
\end{align}
and we consequently obtain the power counting for cumulants
\begin{align}
\label{Wn-g}
    W_{n}\sim g^{-n} g^{n-2}_{\eff} \sim g^{-2}_{0} \xi^{\frac{5(n-2)}{2}+n}\, ,
\end{align}
which gives the familiar critical behavior for non-Gaussian fluctuations, $W_{3}\sim \xi^{9/2}, W_{4}\sim \xi^{7}$. 
The power counting for $W_{n}$ works for equilibrium correlators and we shall assume this counting for off-equilibrium situations.

Now, we estimate the parametric behavior of loop corrections, which is simply given by
\begin{align}
\label{correction}
   \ep_{0}\equiv g^{2}_{\eff} Q^{3}_{*}\,, 
\end{align}
where $Q_{*}$ denotes the characteristic loop momentum (see ref.~\cite{sogabe2021positive} for an explicit demonstration of the counting~\eqref{correction}). 
To obtain eq.~\eqref{correction}, we have used the fact that propagators for rescaled fields is counted as order $1$.
The small parameters which bear certain similarity to $\ep_{0}$ have been introduced in refs~\cite{Mukherjee:2015swa,An:2020vri}.

In this paper, 
we shall work in the limit $\ep_{0}\to 0$ and only keep tree-level contribution to the R.H.S. of the evolution equation~\eqref{SE-eq}. 
We then have the following expression in terms of original fluctuating fields (not for the re-scaled fields~(\ref{rescale})): 
\bes
\label{W-eqs}
\begin{align}
\label{W2-eq}
\pd_{t}W_{2}(t;\xv_1,\xv_2) &=
    \lim_{\equt}  
    \le[ \langle (\d F)_{1}(x_{1})\delta n(x_{2})\rangle_{\con}\ri]_{
    \CC_{\vx_{1},\vx_{2}}
    }\,,
\\
\label{W3-eq}
\pd_{t} W_{3}(t;\xv_1,\ldots,\xv_3) &=
    \lim_{\equt}  
    \le[ \left< (\d F)_{1}(x_{1})\, \prod^{3}_{j=2}\delta n(x_{i}) \right>_{\con}+\left< (\d F)_{2}(x_{1}) \prod^{3}_{j=2}\delta n(x_{i}) \right>_{\con}\ri]_{
    \CC_{\vx_{1},\ldots,\vx_{3}}} \,,
    \\
\label{W4-eq}
\pd_{t} W_{4}(t;\xv_1,\ldots,\xv_4) &=
    \lim_{\equt}  
    \le[ \left< (\d F)_{1}(x_{1}) \prod^{4}_{j=2}\delta n(x_{i}) \right>_{\con}+\ldots+\left< (\d F)_{3}(x_{1}) \prod^{4}_{j=2}\delta n(x_{i}) \right>_{\con}\ri]_{
    \CC_{\vx_{1},\ldots,\vx_{4}} 
    }\,,
\end{align}
\ees
where the prescription for taking the equal-time limit is specified below eq.~\eqref{SE-eq}.  
The expansion of $\d F$ in fluctuating field,  $(\delta F)_n$, can be obtained from the variation of $(\sL)_{n+1}$ with respect to $\psi$ or directly by expanding~$F$, e.g., 
\begin{align}
\label{F1}
(\d F)_{1}&=\vnabla\cdot \le[ \g \na (1-c\na^{2})\delta n - 2 \i \l\vnabla\psi\ri]\,,\\
\label{F2}
(\d F)_{2}&=\vnabla\cdot\le[ \l a' \delta n\vnabla\delta n+\l' a \delta n \vnabla(1-c\na^{2})\delta n- 2\i \l' \delta n\vnabla\psi\ri]\, .   
%     \\
%     \label{F3}
% (\d F)_{3}&=\vnabla\cdot\le[ \frac{1}{2}\l a''\, (\delta n)^{2}\vnabla\delta n+ \l' a'\, (\delta n)^{2}\vnabla\delta n+\l'' a (\delta n)^{2}\, \vnabla(1-c\na^{2})\delta n- i \l''\,(\delta n)^{2} \vnabla\psi\ri]\, ,
\end{align}
To evaluate the R.H.S of eq.~\eqref{W-eqs} at tree level, 
we can use Wick theorem and express $\langle\ldots\rangle_{\con}$ in terms of 
propagators:
\begin{align}
\label{Grr}
G^{\rm rr}(x_1,x_2)&\equiv\langle \delta n(x_{1})\, \psi(x_{2})\rangle\,,
\\
\label{Gra}
G^{\rm ra}(x_1,x_2)&\equiv \langle \delta n(x_{1})\, \psi(x_{2})\rangle\, , 
    \qquad
G^{\rm ar}(x_1,x_2)\equiv \langle \psi(x_{1})\, \delta n(x_{2})\rangle\, , 
\end{align}
where the two-point function of $a$-field is zero by causality (see also appendix~\ref{sec:LTE}). 
We shall not present the explicit expression for the resulting (somewhat complicated) equations here, although obtaining them is straightforward. 
Instead, in the next section, 
we shall first assume an additional separation of scale and further simplify the evolution equations.

\subsection{Patches}

%%%%%%%%%%%%%%%%%%%%%%%%%%%%%%%%%%%%%%%%%%%%%%%%%%
%
%
% A sketch of patches
%
%
% \begin{figure*}[t] 
% %
% %
% %
% \includegraphics[ width=0.6\textwidth]{fig_paper/patch_sketch.png}
% %
% %
% %
% \caption{
% \label{fig:patch}  
% An illustration 
% %
% }
% \end{figure*}
%
%
% 
%%%%%%%%%%%%%%%%%%%%%%%%%%%%%%%%%%%%%%%%%%%%%%%%%%
In many situations of interest, 
the typical expansion rate $\o$ and gradient $\vk$ of the background profile $(n(x))_{0}$ is much smaller than the characteristic momentum $q_{*}$ of the fluctuating modes, 
i.e.,
\begin{align}
\label{epsilon-1}
    \ep_{1}\sim \frac{(\o,k)}{q_{*}}\ll 1 \, . 
\end{align}
For example, $q_{*}$ can be estimated by equating the diffusive rate of the fluctuating modes $\g q^{2}_{*}$ with the expansion rate $\o$, yielding $q_{*}\sim \sqrt{\o/\g}$, which is parametrically larger than $\o$~\cite{Akamatsu_2017} (see also refs.~\cite{An:2019csj}). 
Note, $q_{*}$ is associated with the inverse of the Kibble-Zurek length near the critical point~\cite{Mukherjee:2016kyu}, see ref.~\cite{Akamatsu:2018vjr} for an estimation of this scale for  heavy-ion collisions. 
In what follows, we shall assume the limit eq.~\eqref{epsilon-1}, as refs.~\cite{An:2019csj,An:2020vri} did.

When the separation of scale~\eqref{epsilon-1} is satisfied, 
we can divide the whole system at a given time into many patches with typical length $l_{\rm patch}$ satisfying $q^{-1}\ll l_{\rm patch}\ll k^{-1} $. 
In this limit, 
we can neglect the variation of the background fields within each patch and ignore the correlation of fluctuations at different patches. 
In this sense, we may call those patches ``locally homogeneous and causally disconnected.'' 
Then,
it is convenient to introduce the generalized Wigner transform (WT) of the multi-point correlator as well as their inverse~\cite{An:2020vri}:
\begin{align}
\label{eq:Wig}
W_{n}(\bar{\vx};\vq_{1},...,\vq_{n-1}) & = \int_{\yv_1} \cdots \int_{\yv_n} 
W_{n}(\bar{\vx}+\vy_{1},\ldots,\bar{\vx}+\vy_{n})
\delta\left(\frac{\yv_1+\cdots+\yv_n}{n} \right) \prod^{n}_{j=1}\e^{-\i \vq_{j}\cdot{\vy_{j}}} \, ; %%%
\\
\label{eq:WT-inv}
W_{n}(\vx_{1},...,\vx_{n}) & = \int_{\vq_{1}} \cdots \int_{\vq_n} \, 
W_{n}(\bar \xv;\vq_{1},...,\vq_{n-1})\, 
(2\pi)^{3}\d_{\vq_{1}+\ldots+\vq_{n}}\, 
 \prod^{n}_{j=1}\e^{\i \vq_{j}\cdot{\vx_{j}}} \, . %%%
\end{align}
Here, the integral in the momentum space is abbreviated as $ \int_{\qv} \equiv \int \rmd \qv^3/(2\pi)^3$. For $n=2$, eq.~\eqref{eq:Wig} is reduced to the familiar expression for the Wigner function of Gaussian fluctuations. In eq.~\eqref{eq:Wig},
$\bar{\vx}$ labels each patch, to say the position of its center of energy, while $\vq_{i}$ is the wave vector of the fluctuation field inside the patch. 
In each patch,
the Wigner transform corresponds to the Fourier transform inside the patch.
With this understanding, we shall use Wigner transform and Fourier transform interchangeable when we refer to a patch.  
Since $W_{n}$ is invariant under the shift of all momentum $\vq_{i}$ by the same vector, we can impose the requirement $\sum_i \vq_{i}=0$, meaning only $(n-1)$ $\vq$s are non-redundant.
Hereafter, we shall drop $\qv_{n}$ in the argument of $W_{n}$ to simplify our notations and write   $W_{n}(t,\bar \xv;\qv_1,...,\qv_{n-1})\equiv W_{n}(t,\bar \xv;\qv_1,...,\qv_n)$.

We are interested in deriving evolution equation for $W_{n}$ in Fourier space. 
Applying the generalized Wigner transform to eq.~\eqref{W-eqs}, 
its L.H.S. will be replaced by
\begin{align}
\pd_{t}W_{n}(t,\bar{\xv};\vq_{1},\ldots,\vq_{n-1}) = \cdots \,.
\end{align}
The remaining task is to evaluate the Wigner transform of the R.H.S.. 
For this purpose, 
we must consider the Fourier transform of Lagrangian density $(\sL)_{n}$ and $\delta F$ on each patch.

First, 
by the virtue that intra-patch interaction among fluctuating fields is suppressed, 
we can define Lagrangian density $\sL$ on a given patch $\bar{\vx}$ as
\begin{align}
\sL_{{\rm patch}}(\vx=\bar{\vx}+\vy)= \sL[(n(\vx))_{0}\to (n(\bar{\vx}))_{0};\delta n(\vx)\to \delta n(\vy),\psi(\vx)\to \psi(\vy) ]\, . 
\end{align}
In another word, 
for $\vx=\bar{\vx}+\vy$ lives in the patch $\bar{\vx}$, 
we replace the background fields $(n)_{0}(\vx)$ and $(T)_{0}(\vx)$ by $(n)_{0}(\bar{\vx})$ and $ (T)_{0}(\bar{\vx})$, and the fluctuating fields $\delta n(\vx)$ and $ \psi(\vx)$ by $\delta n(\vy)$ and $ \psi(\vy)$, respectively. Next, on each patch, we consider the action in Fourier space. 
For example,
\begin{align}
\label{L2-q}
\int_{\vy}  (\sL_{\rm patch})_{2}(\bar{\vx}+\vy) = \int {\rm d} t \int_{\vq_1,\vq_2}
\le[ -\psi(\vq_{1})\Gamma(\vq_{2}) \delta n(\vq_{2}) + \i \l (\vq_{1}\cdot\vq_{2})\psi(\vq_{1}) \psi(\vq_{2})\ri]
 \delta_{\vq_{1}+\vq_{2}}\,, 
\end{align}
where the damping rate of the fluctuating modes at momentum $\vq$ is given by
\begin{align}
\label{damping}
\G(\vq)= \g\vq^{2}f(q)\, , 
\end{align}
with $f$ defined in eq.~\eqref{f-def}. 
Similarly, 
we can introduce vertex functions $U_{n,m}$ describing non-linear interaction among fluctuating fields from interacting part of the Lagrangian density ($n>2$),
\begin{align}
\label{L-general}
&\int {\rm d} t\int_{\vy}  (\sL_{{\rm patch}})_{n}(\bar{\vx}+\vy)\notag\\
& = \int {\rm d} t \int_{\vq_1,\ldots,\vq_n}
 \sum^{m_{0}(n)}_{m=1}
\frac{1}{m} U_{n,m}(\vq_1,\ldots,\vq_n) \prod^{m}_{i=1} \psi(\vq_{i})\prod^{n}_{j=m+1} \delta n(\vq_{j})\delta_{\qv_1,...,\qv_n}\, ,
\end{align}
where once again $n$ counts the number of fluctuating fields, whereas $m$ counts that of a-fields. 
In eq.~\eqref{L-general}, 
$m_{0}(n)\leq n$ is the maximum number of $a$-fields in $(\sL)_n$, and one can read these numbers from the Lagrangian, e.g., $m_{0}(3)=2$,  $m_{0}(3)=4$. 
By construction, we require $U_{n,m}$ to be symmetric under each permutation over $\le(\qv_{1},\ldots,\qv_{m}\ri)$ and $\le(\qv_{m+1},\ldots,\qv_{n}\ri)$. 
%Explicitly, we have from \eqref{L3-K} and \eqref{L4-K}
In the expression of $U$, the parameters, $\g, \l, \vartheta_{1,2}$, e.t.c., will depend on $(t,\bar{\vx})$, 
but we shall keep this dependence inexplicit. 
Likewise, we will suppress $(t,\bar{\vx})$-dependence in $W_{n}$s from now on. 
Note the Lagrangian density considered in this work is even in spatial gradient, so that the vertex function is invariant under the spatial inversion:
\begin{align}
\label{U-inversion}
    U_{n,m}(\vq_{1},\ldots,\vq_{n})=
    U_{n,m}(-\vq_{1},\ldots,-\vq_{n})\, . 
\end{align}
With eqs.~\eqref{L2-q} and  \eqref{L-general},
we obtain the Fourier transform of $(\delta F)_{n}$ 
\begin{align}
\label{F1-q}
\le(\delta F\ri)_{1}\stackrel{{\rm WT}}{\longrightarrow}\int_{\vq'_{2}} \le[ -\G(\vq'_{2}) \delta n(\vq'_{2})+ 2 \i\l (\vq_{1}\cdot\vq'_{2}) \psi(\vq'_{2}) \ri] \delta_{-\vq_{1}+\vq'_{2}}\,,
\end{align}
and for $n>2$,
\begin{align}
\label{Fn-q}
& \le(\delta F\ri)_{n-1}\stackrel{{\rm WT}}{\longrightarrow}
 \int_{\vq'_{2},\ldots \vq'_{n}} \sum^{m_{0}(n)}_{m=1}\le[
 U_{n,m}(-\vq_{1},\vq'_{2},\ldots,\vq'_{n}) \prod^{m}_{i=2} 
 \psi(\vq'_{i}) 
 \prod^{n}_{j=m+1}\delta n(\vq'_{j})
\ri] \delta_{-\vq_{1}+\vq'_{2}+\ldots+\vq'_{n}}\, .
\end{align}
% \NB{
% In appendix.~, we present examples which confirm that using \eqref{F1-q} gives the results identical to those from brute-force calculation in the limit \eqref{epsilon-1}. 
% }

\subsection{Evolution equations in Fourier space}

We first consider the evolution equation for $W_{2}$ characterizing Gaussian fluctuations. 
The R.H.S of eq.~\eqref{W2-eq}, i.e., 
\begin{align}
\label{W2-RHS}
\lim_{\equt}\le[
\langle  (\delta F)_{1}(t,\vx_{1})\,\delta n(t_{2},\vx_{2})\, \rangle_{\con}\ri]_{\CC_{\vx_{1},\vx_{2}}}
\stackrel{{\rm WT}}{\longrightarrow}
\lim_{\equt}
\le[
\langle  (\delta F)_{1}(t,\vq_{1})\,\delta n(t',\vq_{2})\rangle_{\con}
\ri]_{\CC_{\vq_{1},\vq_{2}}}\,,
\end{align}
can be divided into two parts from the decomposition $(\delta F)_{1}=(\delta F)_{1,0}+(\delta F)_{1,0}$, where $(\delta F)_{1,m}$ with $ m=0,1$ being the $m$-th power of a-field: 
\begin{align}
\le(\delta F\ri)_{1,0}(\vq_{1})
&= -\int_{\vq'_{2}} \G(\vq'_{2}) \delta n(\vq'_{2}) \delta_{-\vq_{1}+\vq'_{2}}\,,\\
\le(\delta F\ri)_{1,1}(\vq_{1})
&= \int_{\vq'_{2}} 2 \i\l (\vq_{1}\cdot\vq'_{2}) \psi(\vq'_{2}) \delta_{-\vq_{1}+\vq'_{2}}\,.
\end{align}
The first part, corresponding to the contribution from $(\d F)_{1,0}$, is given by 
\begin{align}
\label{W21}
\lim_{\equt} \left[\langle (\delta F)_{1,0}(t,\vq_{1}) \delta n(t',\vq_{2})\rangle_{c} \right]_{\mathcal{C}_{\qv_1,\qv_2}}= -2\Gamma(\vq_{1}) W_{2}(\vq_{1})\,.
\end{align}
Turning to the second part coming from $(\delta F)_{1,1}$, 
we have
\begin{align}
\label{W22}
\lim_{\equt}\left[\langle  (\delta F)_{1,1}(t,\vq_{1}) \delta n(t',\vq_{2}) \rangle_{\con}\right]_{\CC_{ \qv_1,\qv_2 }} = 2\l \vq^{2}_{1}\,, 
\end{align}
where we have used the relation, 
\begin{align}
\label{Gra-equ-t}
\lim_{\equt}\le[
\langle \delta n(t,\vx_{1})\, \psi(t',\vx_{2})
\ri]_{\CC_{\xv_1,\xv_2}}
\stackrel{{\rm WT}}{\longrightarrow} \lim_{\equt}  \le[G^{\rm ra}(t,t';\vq_{1})+G^{\rm ar}(t,t';\vq_{1})\ri]=-\i\,.
\end{align}
To get eq.~\eqref{Gra-equ-t}, we first note $G^{\rm ra}(t,t';\vq)$ and $G^{\rm ar}(t,t';\vq)$ should approach a common $\vq$-dependent constant in the limit $t-t'\to 0^{+}$ and $t-t'\to 0^{-}$, and vanish when $t<t'$ and $t>t'$, respectively. 
Then, we determine this constant to be $-\i$ from the expression for $G^{\rm ra}$ and $G^{\rm ra}$ in the static background~\eqref{Gra-static}, i.e.,
\begin{align}
\label{Gra-equal}
    \lim_{t-t'\to 0^{+}}G^{\rm ra}(t,t';\vq_{1})
    =-\i\, , 
    \qquad
    \lim_{t-t'\to 0^{-}}G^{\rm ar}(t,t';\vq_{1})
    =-\i\, .
\end{align}
Combining eqs.~\eqref{W21} and \eqref{W22}, 
we arrive at the evolution equation for $W_{2}$,
\begin{align}
\label{W2}
\pd_{t}\, W_{2}(\vq_{1})=\le[ -\Gamma(\vq_{1})W_{2}(\vq_{1})\ri]_{\sP_{\vq_{1},\vq_{2}}}
+S_{2}(\vq_{1})\,,
\end{align}
where $\sP_{\vq_{n}}$ denotes the permutation around $\le(\vq_{1},\ldots,\vq_{n}\ri)$. Note this is different from the cyclic operation, e.g., (\ref{cyc_def}), in general.
The source term for $W_{2}$ is given by 
\begin{align}
\label{S2}
    S_{2}(\vq_{1})=2\l \vq^{2}_{1}\,.
\end{align}

Before continuing,
a further remark on the treatment of retarded/advanced response function, $G^{\rm ra}$ and $G^{\rm ar}$, at equal time is due here. 
When computing eq.~\eqref{SE-eq},
we evaluate functions symmetric under the permutation of $(t,t')$. Therefore, $G^{\rm ra}(t,t';\vq)$ and $G^{\rm ar}(t,t';\vq)$ should always appear in the combination as $G^{\rm ra}(t,t';\vq)+ G^{\rm ar}(t,t';\vq)$ and the results should not depend on whether $t-t'\to 0^{+}$ or $t-t'\to 0^{-}$. 
As a consequence, 
we can practically implement eq.~\eqref{Gra-equ-t} with the prescription,
\begin{align}
\label{Gra-prescription}
\lim_{\equt}\,G^{\rm ra}(t,t';\vq_{1})=\lim_{\equt} G^{\rm ar}(t,t';\vq_{1})= -\frac{\i}{2}\, . 
\end{align}
% With \eqref{Gra-prescription}, we then have
% \begin{align}
% \label{C-replacement}
%     \lim_{\equt}\langle F(t,\vx_{i})\, \prod^{n}_{j=2}\, \delta n(t',\vx_{j})\rangle_{c}
%     =
%      \lim_{t'\to t}\langle F(t',\vx_{1})\, \prod^{n}_{j=2}\, \delta n(t,\vx_{j})\rangle_{c}\,.
%      \end{align}
% For example, 
% applying \eqref{C-replacement} together with \eqref{Gra-equ-t}, 
% we reproduce R.H.S of \eqref{W22}

% %%%%%%%%%%%%%%%%%%%%%%%%%%%%%%%%%%%%%%%%%%%%%%%%%%
% %
% %
% % Diagrams for source term
% %
% %
% \begin{figure*}[t] 
% %
% %
% %
% \includegraphics[width=0.6\textwidth]{fig_paper/Sn_diagram.png}
% %
% %
% %
% \caption{
% \label{fig:Sn-dia}  
% An illustration of digrams
% %
% }
% \end{figure*}
% %
% %
% % 
% %%%%%%%%%%%%%%%%%%%%%%%%%%%%%%%%%%%%%%%%%%%%%%%%%%

Now, we return to determine the evolution equations for non-Gaussian fluctuations. 
First, 
we observe the following generalization of~eq.~\eqref{W21}:
\begin{align}
\label{WF1-A}
\lim_{\equt}\left[\left< \vnabla \le(\delta F\ri)_{1,0}(t,\vq_{1}) \prod^{n}_{j=2}\delta n(t_{j},\vq_{j})  \right>_{\con}\right]_{\CC_{\vq_{1},\ldots,\vq_{n}}}
  =
  -n\,\Gamma(\vq_{1})\, W_{n}(\vq_{1},\ldots,\vq_{n-1})\, . 
\end{align}
Moreover, due to the causal structure of SK EFT, 
we have the relation (see appendix.~\ref{sec:LTE} for a derivation)
\begin{align}
\label{WF1-B}
\lim_{\equt}\left< \le(\na^{2}\psi \ri)(t',\vx_{1})\prod^{n}_{j=2}\delta n(t,\vx_{j}) \right>_{\con}=0\,, 
\qquad(\text{for } 
n>2)\,,
\end{align}
which implies
\begin{align}
\label{WF1-B0}
\lim_{t'\to t} \left< \vnabla \le(\delta F\ri)_{1,1}(t',\vx_{1})\prod^{n}_{j=2}\delta n(t,\vx_{j})\right>_{\con}=0\,,
\qquad (\text{for } n>2)\, .
\end{align}
As detailed in appendix.~\ref{sec:LTE}, 
eq.~\eqref{WF1-B} is general and holds for arbitrary $\ep_{0}$ and $\ep_{1}$. 
With eqs.~\eqref{WF1-A} and \eqref{WF1-B} at hand, 
the equations for $W_{3}$ and $W_{4}$ together with that for $W_{2}$ can be collectively written as 
\begin{align}
\label{Wn-evo}
\pd_{t}W_{n}(\vq_{1},\ldots,\vq_{n-1})= \le[-\frac{1}{(n-1)!}\G(\vq_{1})W_{n}(\vq_{1},\ldots,\vq_{n-1}) \ri]_{\sP_{\vq_{1},\ldots,\vq_{n}}}
 +S_{n}(\vq_{1},\ldots,\vq_{n})\, ,
% \\
% \label{W4}
%  &\,  \pd_{t}W_{4}(\vq_{1},\vq_{2},\vq_{3})= \,\le(\,-\frac{1}{3!}\,\G(\vq_{1})W_{4}(\vq_{1},\vq_{2},\vq_{3}) \,\ri)_{\sP_{\vq_{1},\ldots,\vq_{4}}}
%  +S_{4}(\vq_{1},\ldots,\vq_{4})
%  \, ,
\end{align}
where the source terms are given by
\begin{align}
\label{S3-gen}
S_{3}
& \equiv 
\lim_{\equt}\le[
  \left< \le(\delta F\ri)_{2}(\vq_{1})\, \prod^{3}_{j=2}\delta n(\vq_{j}) \right>_{c}
    \ri]_{\CC_{\vq_{1},\ldots,\vq_{3}}} \,, 
\\
\label{S4-gen}
S_{4}&=
    \lim_{\equt}
    \le[
\left< \le(\delta F\ri)_{2}(\vq_{1})\, \prod^{4}_{j=2}\delta n(\vq_{j})\right>_{\con}
    +\left< \le(\delta F\ri)_{3}(\vq_{1})\, \prod^{4}_{j=2}\delta n(\vq_{j}) \right>_{\con}
    \ri]_{\CC_{\vq_{1},\ldots,\vq_{4}}}
    \, . 
\end{align}
Upon substituting eq.~\eqref{Fn-q}, we find the relation,
\begin{align}
\label{dF-dn-general}
    &\langle 
    (\delta F)_{n-1}(\vq_{1})\, \prod^{n}_{l=2}\delta n(\vq_{l})
    \rangle_{\con}\notag\\
    &= \sum^{n-1}_{m=1} 
    \le[
    U_{n,m}(-\vq_{1},\ldots,-\vq_{n})\left(-\frac{\i}{2}\right)^{m-1}\prod^{n}_{j=m+1}W_{2}(\vq_{j})
    \ri]_{\sP_{\vq_{2},\ldots,\vq_{n}}} +\cdots 
    \no \\
    &=\sum^{n-1}_{m=1} 
    \le[
    U_{n,m}(\vq_{1},\ldots,\vq_{n})\left(-\frac{\i}{2}\right)^{m-1}\prod^{n}_{j=m+1}W_{2}(\vq_{j})
    \ri]_{\sP_{\vq_{2},\ldots,\vq_{n}}} +\cdots \,,
\end{align}
where $\sP_{\qv_2,\ldots,\qv_n}$ arises from $(n-1)!$ ways to assign $\vq_{2},\ldots,\vq_{n}$ to $-\vq'_{2},\ldots,-\vq'_{n}$ in $(\delta F)_{n}$ and we have used eq.~\eqref{U-inversion}. 
Note eq.~(\ref{dF-dn-general}) can generally include other contribution $\cdots$ coming from loop corrections. Consequently, the source terms at tree-level take the form
\begin{align}
    \label{S3-q}
   S_{3}(\vq_{1},\ldots,\vq_{3})&= \le[
    \sum^{2}_{m=1}U_{3,m}(\vq_{1},\ldots,\vq_{3})\left(-\frac{\i}{2}\right)^{m-1}\prod^{3}_{j=m+1}W_{2}(\vq_{j})\ri]_{\sP_{\vq_{1},\ldots,\vq_{3}}} \,, 
    \\
\label{S4-q}
S_{4}(\vq_{1},\ldots,\vq_{4})
   & =  
\left[ \sum^{4}_{m=1}
    U_{4,m}(\vq_{1},\ldots,\vq_{4})\left(-\frac{\i}{2}\right)^{m-1}\prod_{j=m+1}W_{2}(\vq_{j}) \right.
  \no \\ 
   &\quad + \left.
    \sum^{2}_{m=1}U_{3,m}(\vq_{1},\vq_{2},-\vq_{12})\left(-\frac{\i}{2}\right)^{m-1}
    \le( W_{2}(\vq_{2})\ri)^{1-\delta_{m2}} W_{3}(\vq_{3},\vq_{4}) \right]_{\sP_{\vq_{1},\ldots,\vq_{4}}}
    \, ,  
\end{align}
where the second line of eq.~\eqref{S4-q} is obtained by generalizing eq.~\eqref{dF-dn-general} and where $q_{ij}$ is defined in eq.~\eqref{qij}. 
Note, we may read eqs.~\eqref{S3-q} and \eqref{S4-q} from the diagrammatic representation:
\begin{align}
\label{S3-dia}
    S_{3}(\vq_{1},\ldots,\vq_{3})&=
  \lim_{\equt}\left[ 
      \raisebox{-29pt}{\tikz[thick]{
    \node at (-3mm,0mm) {$\vq_1$};
    \draw  (0mm,0mm)--(5mm,5mm);
    \draw (5mm,5mm)--(10mm,10mm);
    \node at (8mm,4mm) {$\vq_2$};
    \draw (0mm,0mm)--(5mm,-5mm);
    \draw (5mm,-5mm)--(10mm,-10mm);
    \node at (8mm,-4mm) {$\vq_3$};
    \draw[fill] (0,0) circle (.6ex);
   % \node at (-4mm,-4mm) {$U_{3,1}$};
  }}
  \qquad
   +
  \qquad
   \raisebox{-29pt}{\tikz[thick]{
    \node at (-3mm,0mm) {$\vq_1$};
    \draw (0mm,0mm)--(10mm,10mm);
    \node at (8mm,4mm) {$\vq_2$};
    \draw [aux] (0mm,0mm)--(5mm,-5mm);
    \draw  (5mm,-5mm)--(10mm,-10mm);
    \node at (8mm,-4mm) {$\vq_3$};
    \draw[fill] (0,0) circle (.6ex);
  %  \node at (-4mm,-2mm) {$U_{3,2}$};
  }}\qquad \right]
  \, ,
\\
\label{S4-dia}
    S_{4}(\vq_{1},\ldots,\vq_{4})&=
    \lim_{\equt}
    \left[ \raisebox{-29pt}{\tikz[thick]{
    \draw (0mm,0mm)--(10mm,10mm);
    \node at (10mm,6mm) {$\vq_2$};
    \draw (0mm,0mm)--(14mm,0mm);
    \node at (13mm,-2mm) {$\vq_3$};
    \draw (0mm,0mm)--(10mm,-10mm);
    \node at (10mm,-6mm) {$\vq_4$};
    \draw[fill] (0,0) circle (.6ex);
    \node at (-4mm,-2mm) {$\vq_{1}$};
  }}
  \qquad +
     \raisebox{-29pt}{\tikz[thick]{
    \draw (0mm,0mm)--(10mm,10mm);
    \node at (10mm,6mm) {$\vq_2$};
    \draw (0mm,0mm)--(14mm,0mm);
    \node at (13mm,-2mm) {$\vq_3$};
    \draw [aux](0mm,0mm)--(5mm,-5mm);
    \draw (5mm,-5mm)--(10mm,-10mm);
    \node at (10mm,-6mm) {$\vq_4$};
    \draw[fill] (0,0) circle (.6ex);
    \node at (-4mm,-2mm) {$\vq_{1}$};
  }}
  \qquad +
     \raisebox{-29pt}{\tikz[thick]{
    \draw (0mm,0mm)--(10mm,10mm);
    \node at (10mm,6mm) {$\vq_2$};
    \draw [aux] (0mm,0mm)--(7mm,0mm);
      \draw (7mm,0mm)--(14mm,0mm);
    \node at (13mm,-2mm) {$\vq_3$};
     \draw [aux](0mm,0mm)--(5mm,-5mm);
    \draw (5mm,-5mm)--(10mm,-10mm);
    \node at (10mm,-6mm) {$\vq_4$};
    \draw[fill] (0,0) circle (.6ex);
    \node at (-4mm,-2mm) {$\vq_{1}$};
  }} \right.
  \no \\
  &\qquad \left. 
  \qquad +
     \raisebox{-29pt}{\tikz[thick]{
    \draw [aux](0mm,0mm)--(5mm,5mm);
    \draw (5mm,5mm)--(10mm,10mm);
    \node at (10mm,6mm) {$\vq_2$};
    \draw [aux] (0mm,0mm)--(7mm,0mm);
      \draw (7mm,0mm)--(14mm,0mm);
    \node at (13mm,-2mm) {$\vq_3$};
     \draw [aux](0mm,0mm)--(5mm,-5mm);
    \draw (5mm,-5mm)--(10mm,-10mm);
    \node at (10mm,-6mm) {$\vq_4$};
    \draw[fill] (0,0) circle (.6ex);
    \node at (-4mm,-2mm) {$\vq_{1}$};
  }} \qquad +
   \raisebox{-29pt}{\tikz[thick]{
    \draw (0mm,0mm)--(10mm,10mm);
    \node at (10mm,6mm) {$\vq_2$};
    \draw (5mm,-5mm)--(10mm,0mm);
    \node at (13mm,0) {$\vq_3$};
    \draw (0mm,0mm)--(10mm,-10mm);
    \node at (10mm,-6mm) {$\vq_4$};
    \draw[fill] (0,0) circle (.6ex);
    \node at (-4mm,-2mm) {$\vq_{1}$};
    \draw (5mm,-5mm) circle (.6ex);
  }}
  \qquad +
     \raisebox{-29pt}{\tikz[thick]{
    \draw (0mm,0mm)--(10mm,10mm);
    \node at (10mm,6mm) {$\vq_2$};
    \draw (5mm,-5mm)--(10mm,-10mm);
    \node at (13mm,0) {$\vq_3$};
    \draw (5mm,-5mm)--(10mm,0mm);
    \draw [aux] (0,0)--(5mm,-5mm);
    \draw(5mm,-5mm)--(10mm,-10mm);
    \node at (10mm,-6mm) {$\vq_4$};
    \draw[fill] (0,0) circle (.6ex);
    \node at (-4mm,-2mm) {$\vq_{1}$};
    \draw (5mm,-5mm) circle (.6ex);
  }}
  \qquad \right] \, ,
\end{align}
where we use the solid line and solid-wavy line to represent $G^{\rm rr}$ and $G^{\rm ra}$, respectively:
\begin{align}
\label{propagator}
    G^{\rm rr}(\vq)\, : \qquad
   \raisebox{-12pt}{\tikz[thick]{
    \draw (-14mm,0)--(-0mm,0);
    \node at (-7mm,-3mm) {$\vq$};
    }}\, ;
  \qquad\qquad
 G^{\rm ra}(\vq)\,: \qquad
   \raisebox{-12pt}{\tikz[thick]{
    \draw  (-14mm,0)--(-7mm,0);
    \draw [aux](-7mm,0)--(0,0);
    \node at (-7mm,-3mm) {$\vq$};
    }}\, . 
\end{align}
The empty circles in the second line of eq.~\eqref{S4-dia} denote the insertions of $\i I_{3}$ which give a contribution proportional to $W_{3}$. 
Noting we are taking the equal-time limit, 
in eqs.~\eqref{S3-dia} and \eqref{S4-dia}, 
we shall do the following which gives rise to eqs.~\eqref{S3-gen} and \eqref{S4-gen}:
\begin{enumerate}
    \item Replace solid and solid-wavy lines with $W_{2}$ and $-i/2$ (see eq.~\eqref{Gra-prescription}), respectively;
    \item Insert appropriate vertex $U_{n,m}$;
\item Take care of symmetry factors. 
\end{enumerate}

Finally, by substituting the explicit expression for vertex functions, obtained from eqs.~\eqref{L3} and \eqref{L4},
\begin{align}
\label{U}
U_{3,1}(\vq_1,\ldots,\vq_3)&=
-\frac{1}{2} a' \l  \vq^{2}_{1}+\frac{1}{2}\l' \le[
\frac{\vq_1\cdot \vq_{2}}{\barW(\vq_{2})}\ri]_{\sP_{\vq_{2},\vq_{3}}}\,, \\
U_{3,2}(\vq_1,\ldots,\vq_3)
&=-2\i \l' \vq_{1}\cdot \vq_{2} \,, 
\\
 \label{U4}
   U_{4,1}(\vq_1,\ldots,\vq_{4})&=
-\frac{1}{6}\le(a''\l+2a'\l'\ri)\vq^{2}_{1}
+\frac{1}{12}\l'' \le[
\frac{\vq_1\cdot \vq_{2}}{\barW(\vq_{2})}\ri]_{\sP_{\vq_{2},\ldots,\vq_{4}}}\,,
\\
\label{V4}
U_{4,2}(\vq_1,\ldots,\vq_{4})
&= -\i \l'' \vq_{1}\cdot \vq_{2}\notag\\
&\quad +\i\le[
\vartheta_{1}(\vq_{1}\cdot \vq_{3})(\vq_{2}\cdot \vq_{4})
-(\vartheta_{1}+\vartheta_{2})(\vq_{1}\cdot\vq_{2})(\vq_{3}\cdot \vq_{4})
\ri]_{\sP_{\vq_{1},\vq_{2}}}\notag\\&\hspace{200pt} \times  \frac{1}{\barW_{2}(\vq_{3})\barW_{2}(\vq_{4})}\, , 
\\
U_{4,3}(\vq_1,\ldots,\vq_{4})
&= -\vartheta_{2}\le[ (\vq_{1}\cdot\vq_{2})(\vq_{3}\cdot \vq_{4})\ri]_{\sP_{\vq_{1},\ldots,\vq_{3}}}\, \frac{1}{\barW_{2}(\vq_{4})}\, , 
\\
U_{4,4}(\vq_1,\ldots,\vq_{4})
&= \i\frac{\vartheta_{2}}{6}\le[ (\vq_{1}\cdot\vq_{2})(\vq_{3}\cdot \vq_{4})\ri]_{\sP_{\vq_{1},\ldots,\vq_{4}}}\, ,
\end{align}
% where the momenta taken permutations are those of the fields having the same r or a labels.
we determine explicit expression for source terms,
\begin{align}
\label{S3-B}
&S_{3}(\vq_{1},\ldots,\vq_{3})\notag\\
&=\left[ 
-\frac{1}{2} \G(\vq_{1}) a'\, \barW_{2}(\vq_{1}) W_{2}(\vq_{2})W_{2}(\vq_{2})
+ \l' \vq_{1}\cdot\vq_{2} \le(\frac{W_{2}(\vq_{2})}{\barW_{2}(\vq_{2})}-1\ri) W_{2}(\vq_{3})  \right]_{\sP_{\vq_{1},\ldots,\vq_{3}}}\,,
\\
\label{S4-B}
&S_{4}(\vq_{1},\ldots,\vq_{4})\notag\\
&=
\left[ -\frac{1}{6}a''\,\Gamma(\vq_{1})\barW_{2}(\vq_{1})W_{2}(\vq_{2})W_{2}(\vq_{3})W_{2}(\vq_{4}) - \frac{1}{2}a'\,\Gamma(\vq_{1})\barW_{2}(\vq_{1}) W_{2}(\vq_{2})W_{3}(\vq_{3},\vq_{4}) \right. \no \\
&\quad\ \  -\frac{1}{3}\l' a'q^{2}_{1} W_{2}(\vq_{2})W_{2}(\vq_{3})W_{2}(\vq_{4})+\frac{1}{2}\l'\frac{\vq_{1}\cdot (\vq_{3}+\vq_{4})}{\barW_{2}(\vq_{34})} W_{2}(\vq_{2})W_{3}(\vq_{3},\vq_{4})\no \\
&\quad\ \  +\frac{1}{2}\l''\vq_{1}\cdot\vq_{2}\le(\frac{W_{2}(\vq_{2})}{\barW_{2}(\vq_{2})}-1 \ri) W_{2}(\vq_{3}) W_{2}(\vq_{4})+ \frac{1}{2}\l'\vq_{1}\cdot\vq_{2}\le(\frac{W_{2}(\vq_{2})}{\barW(\vq_{2})}-1\ri) W_{3}(\vq_{3},\vq_{4}) \no \\
&\quad\ \ +\vartheta_{1}\le[
    (\vq_{1}\cdot\vq_{3})(\vq_{2}\cdot\vq_{4})-(\vq_{1}\cdot\vq_{2})(\vq_{3}\cdot \vq_{4})\ri] \frac{W_{2}(\vq_{3})W_{2}(\vq_{4})}{\barW_{2}(\vq_{3})\barW_{2}(\vq_{4})}
    \no\\
    %
    % Theta contribution 
    %
    &\quad\ \ \left. +\vartheta_{2}(\vq_{1}\cdot\vq_{2})(\vq_{3}\cdot \vq_{4})
    \le( 
    -\frac{W_{2}(\vq_{3})W_{2}(\vq_{4})}{\barW_{2}(\vq_{3})\barW_{2}(\vq_{4})}
    + \frac{3}{2}\frac{W_{2}(\vq_{4})}{\barW_{2}(\vq_{4})}-\frac{1}{2}
    \ri) \right]_{\sP_{\vq_{1},\ldots,\vq_{4}}}\,,
\end{align}
where we have used the relation 
\begin{align}
    \l \vq^{2}_{1}= \G(\vq_{1})\,\barW_{2}(\vq_{1})\, ,
\end{align}
and where $\barW_{2}$ is the equilibrium Gaussian fluctuation on the patch $\bar{\vx}$; its explicit expression can be read from eq.~\eqref{W-equi} with  understanding that $a, \xi$, e.t.c., depend on $t,\bar{\vx}$ through their dependence on the background profile $(n)_{0}$ and $T$. 
The evolution equations~\eqref{Wn-evo} together with the explicit expression for the source term eqs.~\eqref{S2}, \eqref{S3-B}, and \eqref{S4-B} are the main results of the present section.

The evolution equation for Gaussian hydrodynamic fluctuations has been studied recently using
Hydro-Kinetic~\cite{Akamatsu_2017,Akamatsu:2018vjr} or Hydro+ formalism~\cite{Stephanov:2017ghc}.
Our expression for $n=2$, \eqref{W2} is consistent with these previous results. 
In hydrodynamic regime where the damping rate $\G(\vq)$~\eqref{damping} is reduced to $\g \qv^{2}$, 
our evolution equations agree with those obtained in ref.~\cite{An:2020vri}. 
The above equations also bear a similarity in structure to evolution equations for cumulants of order parameter field in ref.~\cite{Mukherjee:2015swa}.

The source terms for non-Gaussian fluctuations, $S_{n>2}$, depend on the thermodynamic derivatives of inverse susceptibility $a$ and conductivity $\l$ as well as the additional ``stochastic coefficients'' $\vartheta_{1,2}$. 
As evident from eq.~\eqref{W-equi}, 
the equilibrium fluctuations should only depend on the thermodynamic derivatives of $a$, but not on $\l',\l''$, and $\vartheta_{1,2}$. 
Therefore, the latter should not modify the equilibrium expectations. 
Indeed, we can directly verify that those contributions in $S_{3}$ and $S_{4}$ vanish upon replacing $W_{n}$s with the equilibrium value $\barW_{n}$s given in section~\ref{sec:equilibrium}. 
The remaining terms precisely balance the damping term in evolution equation~\eqref{Wn-evo}, indicating the R.H.S of evolution equation vanishes when $W_{n}$ are evaluated at their equilibrium value~\eqref{W-equi} at it should.

Nevertheless, those $\l',\l''$, and $\vartheta_{1,2}$ contributions, in general, are important in away from the equilibrium settings. 
In particular, we demonstrate explicitly for the first time that non-linear coupling among noise fields, coming from $\vartheta_{2}$ terms in eq.~\eqref{S4-B}, will generate off-equilibrium quadratic fluctuations. 
This is also consistent with the generalized fluctuation-dissipation theorem which relates the real-time $W_{4}$ to four-point retarded correlator~\cite{Wang:1998wg}; the latter depends on $\vartheta_{1,2}$~\cite{Jain:2020fsm}.
As a prior, $\vartheta_{1,2}$ effects may be of importance in the scaling regime $|\qv|\xi \geq 1$, depending on the critical behavior of ``stochastic coefficients'' yet to be explored.

\section{The solution
\label{sec:sol}
}

We are interested in obtaining the solutions to the evolution equation~\eqref{Wn-evo} with generic initial conditions at initial time $t_{\rm I}$:
\begin{align}
\label{WnI}
    W_{n}(\vq_{1},\ldots,\vq_{n-1})\Big|_{t=t_I}=W^{\rm I}_{n}(\vq_{1},\ldots,\vq_{n-1})\, .
\end{align}
In the previous section, we have derived the explicit expression for $S_{n}$ (see eqs.(\ref{S2}), (\ref{S3-q}), and (\ref{S4-q})).
For $n=2$, the evolution equation of the form \eqref{W2} has been studied on semi-realistic hydrodynamic grounds based on Hydro+ framework in refs.~\cite{Du:2020bxp,Rajagopal:2019xwg}. 
In principle, one can do the same for non-Gaussian fluctuations. 
Instead, we shall look for the closed-form solutions to eq.~\eqref{Wn-evo}.

Let us first define the response function $\sG(t,t';\vq)$ which is the solution to the initial value problem,
\begin{align}
\label{G-define}
    \pd_{t}\sG(t,t';\vq)= -\G(\vq)\, \sG(t,t';\vq)\, , 
    \qquad
    \sG(t',t';\vq)=1\, . 
\end{align}
As before, $\G$ as well as $\sG$ depend on the label of the patch $\bar{\xv}$, 
but we will keep this dependence inexplicitly.
In essence, $\sG$ obeys the same equation as linearized equation for the diffusive mode, and is the Green's function describing the diffusion of the density fluctuation. Therefore, one should not be surprised that $\sG$ is proportional to the EFT propagator $G^{\rm ra}$, \eqref{Gra-t} given below. %Moreover, $\sG$ corresponds to a common homogenious solution of the $n$-point evolution equation (\ref{Wn-evo}), in the absence of the source terms, $S_n=0$. 
Then, one can easily verify that the solutions to eq.~\eqref{Wn-evo} with the initial condition~\eqref{WnI} is given by
\begin{align}
\label{Wn-sol}
& W_{n}(t;\vq_{1},\ldots,\vq_{n-1})\notag\\
& =\prod^{n}_{i=1} \sG(t,\tI;\vq_{i}) W^{\rm I}_{n}(\vq_{1},\ldots,\vq_{n-1})+\int^{t}_{\tI} {\rm d} t' \prod^{n}_{j=1}\sG(t,t';\vq_{j}) S_{n}(t';\vq_{1},\ldots,\vq_{n}) \,. 
\end{align}
The first term in eq.~\eqref{W-def} describes the decay of initial fluctuations while the second term represents the dissipation of the fluctuations generated during the evolution of the system.  
Equation~\eqref{Wn-sol} can be implemented in realistic simulation in heavy-ion collisions by 1) computing the response function~(\ref{G-define}) on top of the given hydrodynamic background; 2) evaluating $S_{n}$,~\eqref{S2}, \eqref{S3-B}, and \eqref{S4-B}; 3) performing the integration on time.

\subsection{Field theory interpretation}

From the point of view of field theory, 
the solution (of Schwinger-Dyson equations), eq.~\eqref{Wn-sol} reflects the fact that higher-point functions should be expressible in terms of vertex functions characterizing non-linearity and two-point functions. 
To make this point more transparent, 
we take $W_{3}$ as an example and show explicitly that eq.~\eqref{Wn-sol} can be obtained by directly evaluating tree-level diagrams contributing to $W_{3}$. 
%One can generalize into the higher order correlators $W_{n>3}$, similarly. 

Indeed, 
the cubic correlator $W_{3}$ at $t$ is given by
\begin{align}
\label{W3-diga}
    &W_{3}(\vq_{1},\ldots,\vq_{3})\notag\\
    &= \qquad 
      \raisebox{-29pt}{\tikz[thick]{
    \draw  (-10mm,0)--(-6mm,0);
    \draw [aux] (-6mm,0)--(0,0);
    \node at (-5mm,-3mm) {$\vq_1$};
    \draw (0mm,0mm)--(10mm,10mm);
    \node at (8mm,4mm) {$\vq_2$};
    \draw (0mm,0mm)--(10mm,-10mm);
    \node at (8mm,-4mm) {$\vq_3$};
    \draw[fill] (0,0) circle (.6ex);
    %\node at (0,-8mm) {$(-a')$};
  }}
  \qquad +\qquad
  \raisebox{-29pt}{\tikz[thick]{
    \draw  (-10mm,0)--(-6mm,0);
    \draw [aux] (-6mm,0)--(0,0);
    \node at (-5mm,-3mm) {$\vq_1$};
    \draw [aux] (0mm,0mm)--(5mm,5mm);
    \draw (5mm,5mm)--(10mm,10mm);
    \node at (8mm,4mm) {$\vq_2$};
    \draw (0mm,0mm)--(10mm,-10mm);
    \node at (8mm,-4mm) {$\vq_3$};
    \draw[fill] (0,0) circle (.6ex);
    %\node at (0,-8mm) {$(-a')$};
  }}
    \no \\
    &=\i
    \int^{\infty}_{t_{\rm I}} {\rm d}t' \left[ U_{3,1}(\vq_{1},\ldots,\vq_{3})\, G^{\rm ra}(t,t';\vq_{1})G^{\rm rr}(t,t';\vq_{2})G^{\rm rr}(t,t';\vq_{3})\right.
    \no \\
    &\left.\hspace{70
pt}+ U_{3,2}(\vq_{1},\ldots,\vq_{3})G^{\rm ra}(t,t';\vq_{1})G^{\rm ra}(t,t';\vq_{2})G^{\rm rr}(t,t';\vq_{3}) \right]_{\sP_{\vq_{1},\ldots,\vq_{3}}}\, ,
\end{align}
where $U_{m,n}$ is evaluated at $t'$ and we have used the Feynman rule~\eqref{propagator}. 
Note the difference between diagrams for equilibrium one $\bar W_{3}$ in~eq.~\eqref{W3eq-in-diam} and non-equilibrium $W_{3}$ in eq.~\eqref{W3-diga}. 
Now substituting the expression for propagators (see appendix.~\ref{sec:G-t} for a derivation)
\begin{align}
\label{Gra-t}
    G^{\rm ra}(t,t';\vq)&=-\i\, \sG(t,t';\vq)\theta(t,t')\, , 
    \\
\label{Grr-t}    
    G^{\rm rr}(t,t';\vq) 
    &= \sG(t,t';\vq)W_{2}(t';\vq)\, ,
    \qquad t>t'\,,
\end{align}
into eq.~\eqref{W3-diga} , 
we obtain 
\begin{align}
\label{W3-sol}
&\,    W_{3}(t;\vq_{1},\ldots,\vq_{n-1})=
    \int^{t}_{\tI} {\rm d}t'\, S_{3}(t';\vq_{1},\ldots,\vq_{n-1})\, \prod^{3}_{j=1}\sG(t,t';\vq_{j})\, ,
\end{align}
with $S_{3}$ given by eq.~\eqref{S3-q}. Equation~\eqref{W3-sol} is nothing but the second term on eq.~\eqref{Wn-sol} for $n=3$, as we advertised earlier. 
To recover the first term in eq.~\eqref{Wn-sol}, 
one needs to include an additional boundary term at $t=t_{\rm I}$ to the action by generalizing the method described in ref.~\cite{initial-slip}.

% Our discussion here is partly inspired by the formulation of KMP\"OST framework~\cite{}.  

\section{Summary and outlook
\label{sec:summary}
}

In the view that non-Gaussian fluctuations of baryon density would lead to the important observational signature on the presence of the conjectured QCD critical point,
we have studied the multi-point fluctuations of a generic conserved density based on novel EFT formalism for hydrodynamic fluctuations. 
We obtain not only evolution equations for multi-point equal-time correlators of density fluctuation (see eqs.~\eqref{Wn-evo}, \eqref{S2}, \eqref{S3-q}, and \eqref{S4-q}), which is related to the Schwinger-Dyson equation in the field theory language,  
but also their solutions expressed in terms of the response function (propagator)~\eqref{Wn-sol} for arbitrary initial conditions. 
To the best of our knowledge, the present work is the first one that applies the EFT formalism to study non-Gaussian fluctuations of a diffusive mode near a critical point. 
Our work demonstrates that this EFT approach can be utilized to investigate this intriguing problem systematically based on familiar field theoretical methods.

While the traditional stochastic hydrodynamics only includes Gaussian white noise,
we find non-linear interactions among noise fields, parametrized by (new) ``stochastic coefficients'' $\vartheta_{1,2}$ in eq.~\eqref{L-EFT}~\cite{Jain:2020fsm}, contribute additional to quartic fluctuations ($n=4$) when the system is out of equilibrium. 
Given their potential importance, it would be interesting and instructive to understand the critical behavior of those coefficients in general and in QCD system specifically.

We focus on the pure diffusive mode and does not account for the full interaction among hydrodynamic modes. 
We hope that the main features we obtain here will survive in future more comprehensive studies. 
This anticipation's validity should be checked with the full hydrodynamic EFT theory as shown in refs.~\cite{Crossley:2015,Crossley:2017}.
That said, 
in the view toward providing quantitative and benchmark guidance for the BESII data anticipated in the upcoming years, 
we believe that incorporating the present results, including the compact form solution to the evolution equation, \eqref{Wn-sol} in realistic modelings, as was done previously for Gaussian fluctuations in refs.~\cite{Rajagopal:2019xwg,Du:2020bxp}, is a more immediate priority.

For the conjectured QCD critical point, 
the slowest hydrodynamic mode is entropy per baryon density, $m\equiv s/n$~\cite{Stephanov:2017ghc,Akamatsu:2018vjr}. 
It is also shown in ref.~\cite{Akamatsu:2018vjr} that the Gaussian fluctuation of this mode $\delta m$ will always diverge with the largest Ising critical exponent $\g \approx 1.23$. 
We therefore expect that one should use evolution equation~\eqref{Wn-evo} as well as the solution eq.~\eqref{Wn-sol} with the replacement for the QCD critical point (see ref.~\cite{An:2020vri}):
\begin{align}
    \delta n \to \delta m\, ,
\qquad
a\to \frac{n^{2}}{c_{p}}\, , 
\qquad
\l \to \kappa\,,
\end{align}
where $c_{p}$ is the heat capacity at constant pressure and $\kappa$ is thermal conductivity. 
In a realistic situation, other refinements are needed. For example, the interpretation of $t$ in those equations should be elaborated~\cite{An:2019csj} and might be related to the affine parameter parametrizing the streamline in the background hydrodynamic solution. 
To make contact with experimental observations, a freezeout prescription for $W_{n}$s has to be formulated~\cite{Pradeep:2021opj} (see also refs.~\cite{Oliinychenko:2019zfk,Oliinychenko:2020cmr}).

To obtain the evolution equation~\eqref{Wn-evo}, we have assumed the smallness of two expansion parameter $\epsilon_{0}$ and $\ep_{1}$, defined in eqs.~\eqref{correction} and \eqref{epsilon-1}, respectively. 
The former controls the loop corrections to the present tree-level results, while the latter, if is small, justifies dividing the whole system into a collection of locally homogeneous patches. 
For given hydrodynamic background, whether $\ep_{1}$ is small or not can be checked directly , along the line of section IV. (E) in ref.~\cite{Du:2020bxp}. 
Even in the case that $\ep_{1}$ is not small, one can return to the evolution equation in real space~\eqref{W-eqs} which is valid for arbitrary $\ep_{1}$. 
The importance of the loop corrections  $\ep_{0}$ also needs to be examined, which we defer to future work.

\acknowledgments
We thank Pak Hang Chris Lau and Misha Stephanov for useful discussions.
This work is supported by the Strategic Priority Research Program of Chinese Academy of Sciences, Grant No. XDB34000000.

\appendix
\section{Causal structure
\label{sec:LTE}
}

The causality imposes important constraints on the correaltors in SK formalism. 
In this appendix, we review some of those constraints which we have used at various places in the present work. 

Let us begin with the largest time equation (LTE) which, in the context of this work, states that for multi-point correlator of the fluctuating fields, denoted collectively by $\phi_{s}$ $( s={\rm a},{\rm r})$,
must satisfy
\begin{align}
\label{LTE}
G_{s_{1},\ldots,s_{n},s_{n+1}}(t_{1},\ldots,t_{n},t_{n+1})  = 
\langle \phi_{s_{1}}(t_{1})\,\ldots \phi_{s_{n}}(t_{n})\phi_{\rm a}(t_{n+1})\rangle_{\con}=0,
\qquad
\,t_{n+1}> t_{1},\ldots,t_{n}
\, . 
\end{align}
In this section, we shall suppress spatial dependence in our notations whenever possible, as this is sufficient for the present purpose of studying causal structure. 
Physically, LTE~\eqref{LTE} tells us that the physical observables (r-field) can only be correlated with noise (a-field) in the past but not in future. 
Applying LTE, 
we recover the familiar results $G^{\rm ra}(t',t)=0$ if $t'<t$ and $G^{\rm aa}$ is identically zero. 
The interested reader can consult ref.~\cite{Gao:2018bxz} for a proof of LTE in general situations.

A related property is that any diagrams that contains a loop made entirely of $G^{\rm ar}$ propagator should vanish. 
Let us denote the vertices in such a loop by $V_{1},\ldots, V_{n}$ which are inserted at $t_{1},\ldots,t_{n}$, respectively. 
Since this loop only contains $G^{\rm ar}(t,t')\propto \theta(t'-t)$, 
it is proportional to 
\begin{align}
\label{ar-int}
    %\int d_{1}\ldots \int d t_{n}\,
    \propto \theta(t_2-t_{1})\, \theta(t_3-t_{2})\, \ldots \theta(t_{n}-t_{n-1})\, \theta(t_{1}-t_{n}) =0\, . 
\end{align}
The first $n-1$ step functions will vanish unless $t_{n}>t_{n-1}\ldots > t_{1}$. 
However, the last step function is non-zero only if $t_{1}>t_{n}$, 
implying eq.~\eqref{ar-int}. 
For example, the loop diagram below vanishes:
\begin{align}
\label{loop-zero}
    \raisebox{-29pt}{\tikz[thick]{ 
    \node at (7mm,0mm) {$t$};
        \draw  (12mm,0) arc (180:90:6mm);
        \draw [aux] (18mm,6mm) arc
        (90:0:6mm);
        \draw [aux] (12mm,0) arc (180:270:6mm); \draw (18mm,-6mm) arc
        (-90:0:6mm); 
        \node at (29mm,-0mm){$t'$};
      }}
\end{align}

Now, let us prove eq.~\eqref{WF1-B}. 
We shall first show:
\begin{align}
\label{n-a-zero}
\lim_{t\to t'}\langle  \psi(t,\vx_{1}) \prod^{n}_{j=2}\delta n(t',\vx_{j}) \rangle_{\con}=0\, , 
\qquad
(\text{for }n>2)\, . 
\end{align}
Our proof is inspired by the method used in ref.~\cite{Gao:2018bxz}. 
Note LTE~\eqref{LTE} implies that eq.~\eqref{n-a-zero} vanishes for $t-t'\to 0^{+}$. 
In what follow, we shall consider the case, $t-t'\to 0^{-}$, from now on.

First, we note that \eqref{n-a-zero} is zero for $n>2$ if we only include Gaussian part of the action,
so a non-zero contribution to eq.~\eqref{n-a-zero} must involve the insertion of vertices.
Let us consider the external a-field $\psi(t)$ 
which can only connect to a r-leg of a vertex $V_{1}$ at time $t_{1}$. 
Since all vertices must contain at least one a-leg,
we can continue following an a-leg in $V_{1}$ which in turn has to connect with a r-leg through a $G^{ar}$ at the second vertex $V_{2}$ at time $t_{2}$. 
Repeating this procedure, we find two possibilities (see also ref.~\cite{Gao:2018bxz}):
\begin{enumerate}
    \item 
    We end up with an internal vertex $V_{i}$ at $t_{i}$ we have already encountered, meaning we have an internal loop formed solely by $G^{ar}$. 
    As we explained earlier, the whole diagram is zero. 

\item We come to the external vertex $V_{0}$ at $t_{0}$ which connect with one r-field $\delta n(t)$ through a-leg. 
In this case, the diagram is proportional to
\begin{align}
       \int {\rm d} t_{0} {\rm d} t_{1}\ldots {\rm d} t_{n} \theta(t_1-t)  \ldots \theta(t'-t_{0})\,,
\end{align}
which also vanishes in the limit $t\to t'$. 
For example, 
\begin{align}
\lim_{t\to t'}
\raisebox{-29pt}{\tikz[thick]{
\node at (-13mm,0mm) {$t$};
    \draw [aux] (-10mm,0)--(-4mm,0);
    \draw  (-4mm,0)--(0,0);
    \draw [aux](0mm,0mm)--(5mm,5mm);
     \draw (5mm,5mm)--(10mm,10mm);
    \node at (10mm,12mm) {$t'$};
    \draw (0mm,0mm)--(10mm,-10mm);
    \node at (0mm,-4mm) {$t_{0}$};
    \draw[fill] (0,0) circle (.6ex);
  }}
  \qquad
  \propto \lim_{t\to t'}\int^{t'}_{t} {\rm d} t_{0} =0\,.
\end{align} 
\end{enumerate}

The above analysis is unchanged when arbitrary power of spatial derivatives acting on eq.~\eqref{n-a-zero}.
We conclude that eq.~\eqref{WF1-B} holds generically due to the causal structure of SK EFT.

\section{The evolution equations for un-equal time correlators
\label{sec:G-t}
}

In the main body of the present work, we focus on the equal-time correlators of density fluctuations. 
Our method can be readily generalized to study un-equal time correlators, such as $G^{\rm rr}(x,x')$ defined in eq.~\eqref{Grr}.
From Schwinger-Dyson equation~\eqref{dnn},  we have:
\begin{align}
    \pd_{t}G^{\rm rr}(x,x')= \langle \delta F(x)\, \delta n(x')\rangle\, . 
\end{align}
Assuming $\ep_{0}$ is small, we could keep the tree level contribution which comes from $(\delta F)_{1}$:
\begin{align}
\label{dGrr}
    \pd_{t}G^{\rm rr}(x,x')= \langle (\delta F)_{1}(x)\, \delta n (x')\rangle\, , 
    \qquad 
    %\text{for}\, 
    t>t'\, . 
\end{align}
Without losing generality, we assume $t>t'$ and consequently $(\delta F)_{1,1}$ will not contribute because of LTE~\eqref{LTE}. 
Further consider the limit $\ep_{1}\to 0$ and apply the Wigner transform to eq.~\eqref{dGrr}, 
we find
\begin{align}
\label{dGrr-q}
    \pd_{t}G^{\rm rr}(t,t';\vq)= -\G(\vq)\, G^{\rm rr}(t,t';\vq)\, , 
    \qquad
    t>t'\, . 
\end{align}
Solving eq.~\eqref{dGrr-q} with the initial condition $G(t,t;\vq)=W_{2}(t;\vq)$, 
we obtain eq.~\eqref{Grr-t}. 
Following a similar procedure,
we have the equation for $G^{\rm ra}$,
\begin{align}
\label{dGra-q}
    \pd_{t}G^{\rm ra}(t,t';\vq)= -\G(\vq)\, G^{\rm ra}(t,t';\vq)\, , 
    \qquad
    t>t'\, . 
\end{align}
with the initial condition eq.~\eqref{Gra-equ-t}.
This leads to eq.~\eqref{Gra-t}. 

We can test eq.~\eqref{Gra-equ-t} by considering the static background.
In this situation, the tree-level propagators can be read from eq.~\eqref{L2}:
\begin{align}
\label{Gra-static}
G^{\rm ra}(t,t';\qv) &= -\i\theta(t-t')\, \e^{-\Gamma(\vq)(t-t')}\,, 
\qquad G^{\rm ar}(t,t';\qv) = -\i\theta(t'-t)\, \e^{-\Gamma(\vq)(t-t')}\,, 
\end{align}
which agrees with eq.~\eqref{Gra-equ-t}.

\bibliographystyle{apsrev4-1}
\bibliography{refs.bib}

\begin{thebibliography}{45}%
\makeatletter
\providecommand \@ifxundefined [1]{%
 \@ifx{#1\undefined}
}%
\providecommand \@ifnum [1]{%
 \ifnum #1\expandafter \@firstoftwo
 \else \expandafter \@secondoftwo
 \fi
}%
\providecommand \@ifx [1]{%
 \ifx #1\expandafter \@firstoftwo
 \else \expandafter \@secondoftwo
 \fi
}%
\providecommand \natexlab [1]{#1}%
\providecommand \enquote  [1]{``#1''}%
\providecommand \bibnamefont  [1]{#1}%
\providecommand \bibfnamefont [1]{#1}%
\providecommand \citenamefont [1]{#1}%
\providecommand \href@noop [0]{\@secondoftwo}%
\providecommand \href [0]{\begingroup \@sanitize@url \@href}%
\providecommand \@href[1]{\@@startlink{#1}\@@href}%
\providecommand \@@href[1]{\endgroup#1\@@endlink}%
\providecommand \@sanitize@url [0]{\catcode `\\12\catcode `\$12\catcode
  `\&12\catcode `\#12\catcode `\^12\catcode `\_12\catcode `\%12\relax}%
\providecommand \@@startlink[1]{}%
\providecommand \@@endlink[0]{}%
\providecommand \url  [0]{\begingroup\@sanitize@url \@url }%
\providecommand \@url [1]{\endgroup\@href {#1}{\urlprefix }}%
\providecommand \urlprefix  [0]{URL }%
\providecommand \Eprint [0]{\href }%
\providecommand \doibase [0]{http://dx.doi.org/}%
\providecommand \selectlanguage [0]{\@gobble}%
\providecommand \bibinfo  [0]{\@secondoftwo}%
\providecommand \bibfield  [0]{\@secondoftwo}%
\providecommand \translation [1]{[#1]}%
\providecommand \BibitemOpen [0]{}%
\providecommand \bibitemStop [0]{}%
\providecommand \bibitemNoStop [0]{.\EOS\space}%
\providecommand \EOS [0]{\spacefactor3000\relax}%
\providecommand \BibitemShut  [1]{\csname bibitem#1\endcsname}%
\let\auto@bib@innerbib\@empty
%</preamble>
\bibitem [{\citenamefont {Stephanov}\ \emph {et~al.}(1998)\citenamefont
  {Stephanov}, \citenamefont {Rajagopal},\ and\ \citenamefont
  {Shuryak}}]{Stephanov:1998dy}%
  \BibitemOpen
  \bibfield  {author} {\bibinfo {author} {\bibfnamefont {M.~A.}\ \bibnamefont
  {Stephanov}}, \bibinfo {author} {\bibfnamefont {K.}~\bibnamefont
  {Rajagopal}}, \ and\ \bibinfo {author} {\bibfnamefont {E.~V.}\ \bibnamefont
  {Shuryak}},\ }\href {\doibase 10.1103/PhysRevLett.81.4816} {\bibfield
  {journal} {\bibinfo  {journal} {Phys. Rev. Lett.}\ }\textbf {\bibinfo
  {volume} {81}},\ \bibinfo {pages} {4816} (\bibinfo {year} {1998})},\ \Eprint
  {http://arxiv.org/abs/hep-ph/9806219} {arXiv:hep-ph/9806219} \BibitemShut
  {NoStop}%
\bibitem [{\citenamefont {Stephanov}\ \emph {et~al.}(1999)\citenamefont
  {Stephanov}, \citenamefont {Rajagopal},\ and\ \citenamefont
  {Shuryak}}]{Stephanov:1999zu}%
  \BibitemOpen
  \bibfield  {author} {\bibinfo {author} {\bibfnamefont {M.~A.}\ \bibnamefont
  {Stephanov}}, \bibinfo {author} {\bibfnamefont {K.}~\bibnamefont
  {Rajagopal}}, \ and\ \bibinfo {author} {\bibfnamefont {E.~V.}\ \bibnamefont
  {Shuryak}},\ }\href {\doibase 10.1103/PhysRevD.60.114028} {\bibfield
  {journal} {\bibinfo  {journal} {Phys. Rev. D}\ }\textbf {\bibinfo {volume}
  {60}},\ \bibinfo {pages} {114028} (\bibinfo {year} {1999})},\ \Eprint
  {http://arxiv.org/abs/hep-ph/9903292} {arXiv:hep-ph/9903292} \BibitemShut
  {NoStop}%
\bibitem [{\citenamefont {Aprahamian}\ \emph {et~al.}(2015)\citenamefont
  {Aprahamian} \emph {et~al.}}]{Aprahamian:2015qub}%
  \BibitemOpen
  \bibfield  {author} {\bibinfo {author} {\bibfnamefont {A.}~\bibnamefont
  {Aprahamian}} \emph {et~al.},\ }\href@noop {} {\  (\bibinfo {year}
  {2015})}\BibitemShut {NoStop}%
\bibitem [{\citenamefont {Busza}\ \emph {et~al.}(2018)\citenamefont {Busza},
  \citenamefont {Rajagopal},\ and\ \citenamefont {van~der
  Schee}}]{Busza:2018rrf}%
  \BibitemOpen
  \bibfield  {author} {\bibinfo {author} {\bibfnamefont {W.}~\bibnamefont
  {Busza}}, \bibinfo {author} {\bibfnamefont {K.}~\bibnamefont {Rajagopal}}, \
  and\ \bibinfo {author} {\bibfnamefont {W.}~\bibnamefont {van~der Schee}},\
  }\href {\doibase 10.1146/annurev-nucl-101917-020852} {\bibfield  {journal}
  {\bibinfo  {journal} {Ann. Rev. Nucl. Part. Sci.}\ }\textbf {\bibinfo
  {volume} {68}},\ \bibinfo {pages} {339} (\bibinfo {year} {2018})},\ \Eprint
  {http://arxiv.org/abs/1802.04801} {arXiv:1802.04801 [hep-ph]} \BibitemShut
  {NoStop}%
\bibitem [{\citenamefont {Bzdak}\ \emph {et~al.}(2020)\citenamefont {Bzdak},
  \citenamefont {Esumi}, \citenamefont {Koch}, \citenamefont {Liao},
  \citenamefont {Stephanov},\ and\ \citenamefont {Xu}}]{Bzdak:2019pkr}%
  \BibitemOpen
  \bibfield  {author} {\bibinfo {author} {\bibfnamefont {A.}~\bibnamefont
  {Bzdak}}, \bibinfo {author} {\bibfnamefont {S.}~\bibnamefont {Esumi}},
  \bibinfo {author} {\bibfnamefont {V.}~\bibnamefont {Koch}}, \bibinfo {author}
  {\bibfnamefont {J.}~\bibnamefont {Liao}}, \bibinfo {author} {\bibfnamefont
  {M.}~\bibnamefont {Stephanov}}, \ and\ \bibinfo {author} {\bibfnamefont
  {N.}~\bibnamefont {Xu}},\ }\href {\doibase 10.1016/j.physrep.2020.01.005}
  {\bibfield  {journal} {\bibinfo  {journal} {Phys. Rept.}\ }\textbf {\bibinfo
  {volume} {853}},\ \bibinfo {pages} {1} (\bibinfo {year} {2020})},\ \Eprint
  {http://arxiv.org/abs/1906.00936} {arXiv:1906.00936 [nucl-th]} \BibitemShut
  {NoStop}%
\bibitem [{\citenamefont {Fukushima}\ and\ \citenamefont
  {Hatsuda}(2011)}]{Fukushima:2010bq}%
  \BibitemOpen
  \bibfield  {author} {\bibinfo {author} {\bibfnamefont {K.}~\bibnamefont
  {Fukushima}}\ and\ \bibinfo {author} {\bibfnamefont {T.}~\bibnamefont
  {Hatsuda}},\ }\href {\doibase 10.1088/0034-4885/74/1/014001} {\bibfield
  {journal} {\bibinfo  {journal} {Rept. Prog. Phys.}\ }\textbf {\bibinfo
  {volume} {74}},\ \bibinfo {pages} {014001} (\bibinfo {year} {2011})},\
  \Eprint {http://arxiv.org/abs/1005.4814} {arXiv:1005.4814 [hep-ph]}
  \BibitemShut {NoStop}%
\bibitem [{\citenamefont {Ikeda}\ \emph {et~al.}(2021)\citenamefont {Ikeda},
  \citenamefont {Kharzeev},\ and\ \citenamefont {Kikuchi}}]{Ikeda:2020agk}%
  \BibitemOpen
  \bibfield  {author} {\bibinfo {author} {\bibfnamefont {K.}~\bibnamefont
  {Ikeda}}, \bibinfo {author} {\bibfnamefont {D.~E.}\ \bibnamefont {Kharzeev}},
  \ and\ \bibinfo {author} {\bibfnamefont {Y.}~\bibnamefont {Kikuchi}},\ }\href
  {\doibase 10.1103/PhysRevD.103.L071502} {\bibfield  {journal} {\bibinfo
  {journal} {Phys. Rev. D}\ }\textbf {\bibinfo {volume} {103}},\ \bibinfo
  {pages} {L071502} (\bibinfo {year} {2021})},\ \Eprint
  {http://arxiv.org/abs/2012.02926} {arXiv:2012.02926 [hep-ph]} \BibitemShut
  {NoStop}%
\bibitem [{\citenamefont {Honda}\ \emph {et~al.}(2021)\citenamefont {Honda},
  \citenamefont {Itou}, \citenamefont {Kikuchi},\ and\ \citenamefont
  {Tanizaki}}]{Honda:2021ovk}%
  \BibitemOpen
  \bibfield  {author} {\bibinfo {author} {\bibfnamefont {M.}~\bibnamefont
  {Honda}}, \bibinfo {author} {\bibfnamefont {E.}~\bibnamefont {Itou}},
  \bibinfo {author} {\bibfnamefont {Y.}~\bibnamefont {Kikuchi}}, \ and\
  \bibinfo {author} {\bibfnamefont {Y.}~\bibnamefont {Tanizaki}},\ }\href@noop
  {} {\  (\bibinfo {year} {2021})},\ \Eprint {http://arxiv.org/abs/2110.14105}
  {arXiv:2110.14105 [hep-th]} \BibitemShut {NoStop}%
\bibitem [{\citenamefont {Berdnikov}\ and\ \citenamefont
  {Rajagopal}(2000)}]{Berdnikov:1999ph}%
  \BibitemOpen
  \bibfield  {author} {\bibinfo {author} {\bibfnamefont {B.}~\bibnamefont
  {Berdnikov}}\ and\ \bibinfo {author} {\bibfnamefont {K.}~\bibnamefont
  {Rajagopal}},\ }\href {\doibase 10.1103/PhysRevD.61.105017} {\bibfield
  {journal} {\bibinfo  {journal} {Phys. Rev. D}\ }\textbf {\bibinfo {volume}
  {61}},\ \bibinfo {pages} {105017} (\bibinfo {year} {2000})},\ \Eprint
  {http://arxiv.org/abs/hep-ph/9912274} {arXiv:hep-ph/9912274} \BibitemShut
  {NoStop}%
\bibitem [{\citenamefont {Mukherjee}\ \emph {et~al.}(2015)\citenamefont
  {Mukherjee}, \citenamefont {Venugopalan},\ and\ \citenamefont
  {Yin}}]{Mukherjee:2015swa}%
  \BibitemOpen
  \bibfield  {author} {\bibinfo {author} {\bibfnamefont {S.}~\bibnamefont
  {Mukherjee}}, \bibinfo {author} {\bibfnamefont {R.}~\bibnamefont
  {Venugopalan}}, \ and\ \bibinfo {author} {\bibfnamefont {Y.}~\bibnamefont
  {Yin}},\ }\href {\doibase 10.1103/PhysRevC.92.034912} {\bibfield  {journal}
  {\bibinfo  {journal} {Phys. Rev. C}\ }\textbf {\bibinfo {volume} {92}},\
  \bibinfo {pages} {034912} (\bibinfo {year} {2015})},\ \Eprint
  {http://arxiv.org/abs/1506.00645} {arXiv:1506.00645 [hep-ph]} \BibitemShut
  {NoStop}%
\bibitem [{\citenamefont {Yin}(2018)}]{Yin:2018ejt}%
  \BibitemOpen
  \bibfield  {author} {\bibinfo {author} {\bibfnamefont {Y.}~\bibnamefont
  {Yin}},\ }\href@noop {} {\  (\bibinfo {year} {2018})},\ \Eprint
  {http://arxiv.org/abs/1811.06519} {arXiv:1811.06519 [nucl-th]} \BibitemShut
  {NoStop}%
\bibitem [{\citenamefont {Nahrgang}\ \emph {et~al.}(2019)\citenamefont
  {Nahrgang}, \citenamefont {Bluhm}, \citenamefont {Schaefer},\ and\
  \citenamefont {Bass}}]{Nahrgang:2018afz}%
  \BibitemOpen
  \bibfield  {author} {\bibinfo {author} {\bibfnamefont {M.}~\bibnamefont
  {Nahrgang}}, \bibinfo {author} {\bibfnamefont {M.}~\bibnamefont {Bluhm}},
  \bibinfo {author} {\bibfnamefont {T.}~\bibnamefont {Schaefer}}, \ and\
  \bibinfo {author} {\bibfnamefont {S.~A.}\ \bibnamefont {Bass}},\ }\href
  {\doibase 10.1103/PhysRevD.99.116015} {\bibfield  {journal} {\bibinfo
  {journal} {Phys. Rev. D}\ }\textbf {\bibinfo {volume} {99}},\ \bibinfo
  {pages} {116015} (\bibinfo {year} {2019})},\ \Eprint
  {http://arxiv.org/abs/1804.05728} {arXiv:1804.05728 [nucl-th]} \BibitemShut
  {NoStop}%
\bibitem [{\citenamefont {An}\ \emph {et~al.}(2022)\citenamefont {An} \emph
  {et~al.}}]{An:2021wof}%
  \BibitemOpen
  \bibfield  {author} {\bibinfo {author} {\bibfnamefont {X.}~\bibnamefont {An}}
  \emph {et~al.},\ }\href {\doibase 10.1016/j.nuclphysa.2021.122343} {\bibfield
   {journal} {\bibinfo  {journal} {Nucl. Phys. A}\ }\textbf {\bibinfo {volume}
  {1017}},\ \bibinfo {pages} {122343} (\bibinfo {year} {2022})},\ \Eprint
  {http://arxiv.org/abs/2108.13867} {arXiv:2108.13867 [nucl-th]} \BibitemShut
  {NoStop}%
\bibitem [{\citenamefont {Kapusta}\ \emph {et~al.}(2012)\citenamefont
  {Kapusta}, \citenamefont {Muller},\ and\ \citenamefont
  {Stephanov}}]{Kapusta:2011gt}%
  \BibitemOpen
  \bibfield  {author} {\bibinfo {author} {\bibfnamefont {J.~I.}\ \bibnamefont
  {Kapusta}}, \bibinfo {author} {\bibfnamefont {B.}~\bibnamefont {Muller}}, \
  and\ \bibinfo {author} {\bibfnamefont {M.}~\bibnamefont {Stephanov}},\ }\href
  {\doibase 10.1103/PhysRevC.85.054906} {\bibfield  {journal} {\bibinfo
  {journal} {Phys. Rev. C}\ }\textbf {\bibinfo {volume} {85}},\ \bibinfo
  {pages} {054906} (\bibinfo {year} {2012})},\ \Eprint
  {http://arxiv.org/abs/1112.6405} {arXiv:1112.6405 [nucl-th]} \BibitemShut
  {NoStop}%
\bibitem [{\citenamefont {Akamatsu}\ \emph {et~al.}(2017)\citenamefont
  {Akamatsu}, \citenamefont {Mazeliauskas},\ and\ \citenamefont
  {Teaney}}]{Akamatsu_2017}%
  \BibitemOpen
  \bibfield  {author} {\bibinfo {author} {\bibfnamefont {Y.}~\bibnamefont
  {Akamatsu}}, \bibinfo {author} {\bibfnamefont {A.}~\bibnamefont
  {Mazeliauskas}}, \ and\ \bibinfo {author} {\bibfnamefont {D.}~\bibnamefont
  {Teaney}},\ }\href {\doibase 10.1103/physrevc.95.014909} {\bibfield
  {journal} {\bibinfo  {journal} {Physical Review C}\ }\textbf {\bibinfo
  {volume} {95}} (\bibinfo {year} {2017}),\
  10.1103/physrevc.95.014909}\BibitemShut {NoStop}%
\bibitem [{\citenamefont {Stephanov}\ and\ \citenamefont
  {Yin}(2018)}]{Stephanov:2017ghc}%
  \BibitemOpen
  \bibfield  {author} {\bibinfo {author} {\bibfnamefont {M.}~\bibnamefont
  {Stephanov}}\ and\ \bibinfo {author} {\bibfnamefont {Y.}~\bibnamefont
  {Yin}},\ }\href {\doibase 10.1103/PhysRevD.98.036006} {\bibfield  {journal}
  {\bibinfo  {journal} {Phys. Rev.}\ }\textbf {\bibinfo {volume} {D98}},\
  \bibinfo {pages} {036006} (\bibinfo {year} {2018})},\ \Eprint
  {http://arxiv.org/abs/1712.10305} {arXiv:1712.10305 [nucl-th]} \BibitemShut
  {NoStop}%
%%CITATION = ARXIV:1712.10305;%%
\bibitem [{\citenamefont {Martinez}\ and\ \citenamefont
  {Sch\"afer}(2019)}]{Martinez:2018wia}%
  \BibitemOpen
  \bibfield  {author} {\bibinfo {author} {\bibfnamefont {M.}~\bibnamefont
  {Martinez}}\ and\ \bibinfo {author} {\bibfnamefont {T.}~\bibnamefont
  {Sch\"afer}},\ }\href {\doibase 10.1103/PhysRevC.99.054902} {\bibfield
  {journal} {\bibinfo  {journal} {Phys. Rev. C}\ }\textbf {\bibinfo {volume}
  {99}},\ \bibinfo {pages} {054902} (\bibinfo {year} {2019})},\ \Eprint
  {http://arxiv.org/abs/1812.05279} {arXiv:1812.05279 [hep-th]} \BibitemShut
  {NoStop}%
\bibitem [{\citenamefont {An}\ \emph {et~al.}(2020)\citenamefont {An},
  \citenamefont {Basar}, \citenamefont {Stephanov},\ and\ \citenamefont
  {Yee}}]{An:2019csj}%
  \BibitemOpen
  \bibfield  {author} {\bibinfo {author} {\bibfnamefont {X.}~\bibnamefont
  {An}}, \bibinfo {author} {\bibfnamefont {G.}~\bibnamefont {Basar}}, \bibinfo
  {author} {\bibfnamefont {M.}~\bibnamefont {Stephanov}}, \ and\ \bibinfo
  {author} {\bibfnamefont {H.-U.}\ \bibnamefont {Yee}},\ }\href {\doibase
  10.1103/PhysRevC.102.034901} {\bibfield  {journal} {\bibinfo  {journal}
  {Phys. Rev. C}\ }\textbf {\bibinfo {volume} {102}},\ \bibinfo {pages}
  {034901} (\bibinfo {year} {2020})},\ \Eprint
  {http://arxiv.org/abs/1912.13456} {arXiv:1912.13456 [hep-th]} \BibitemShut
  {NoStop}%
\bibitem [{\citenamefont {Akamatsu}\ \emph {et~al.}(2019)\citenamefont
  {Akamatsu}, \citenamefont {Teaney}, \citenamefont {Yan},\ and\ \citenamefont
  {Yin}}]{Akamatsu:2018vjr}%
  \BibitemOpen
  \bibfield  {author} {\bibinfo {author} {\bibfnamefont {Y.}~\bibnamefont
  {Akamatsu}}, \bibinfo {author} {\bibfnamefont {D.}~\bibnamefont {Teaney}},
  \bibinfo {author} {\bibfnamefont {F.}~\bibnamefont {Yan}}, \ and\ \bibinfo
  {author} {\bibfnamefont {Y.}~\bibnamefont {Yin}},\ }\href {\doibase
  10.1103/PhysRevC.100.044901} {\bibfield  {journal} {\bibinfo  {journal}
  {Phys. Rev. C}\ }\textbf {\bibinfo {volume} {100}},\ \bibinfo {pages}
  {044901} (\bibinfo {year} {2019})},\ \Eprint
  {http://arxiv.org/abs/1811.05081} {arXiv:1811.05081 [nucl-th]} \BibitemShut
  {NoStop}%
\bibitem [{\citenamefont {Rajagopal}\ \emph {et~al.}(2020)\citenamefont
  {Rajagopal}, \citenamefont {Ridgway}, \citenamefont {Weller},\ and\
  \citenamefont {Yin}}]{Rajagopal:2019xwg}%
  \BibitemOpen
  \bibfield  {author} {\bibinfo {author} {\bibfnamefont {K.}~\bibnamefont
  {Rajagopal}}, \bibinfo {author} {\bibfnamefont {G.}~\bibnamefont {Ridgway}},
  \bibinfo {author} {\bibfnamefont {R.}~\bibnamefont {Weller}}, \ and\ \bibinfo
  {author} {\bibfnamefont {Y.}~\bibnamefont {Yin}},\ }\href {\doibase
  10.1103/PhysRevD.102.094025} {\bibfield  {journal} {\bibinfo  {journal}
  {Phys. Rev. D}\ }\textbf {\bibinfo {volume} {102}},\ \bibinfo {pages}
  {094025} (\bibinfo {year} {2020})},\ \Eprint
  {http://arxiv.org/abs/1908.08539} {arXiv:1908.08539 [hep-ph]} \BibitemShut
  {NoStop}%
\bibitem [{\citenamefont {Du}\ \emph {et~al.}(2020)\citenamefont {Du},
  \citenamefont {Heinz}, \citenamefont {Rajagopal},\ and\ \citenamefont
  {Yin}}]{Du:2020bxp}%
  \BibitemOpen
  \bibfield  {author} {\bibinfo {author} {\bibfnamefont {L.}~\bibnamefont
  {Du}}, \bibinfo {author} {\bibfnamefont {U.}~\bibnamefont {Heinz}}, \bibinfo
  {author} {\bibfnamefont {K.}~\bibnamefont {Rajagopal}}, \ and\ \bibinfo
  {author} {\bibfnamefont {Y.}~\bibnamefont {Yin}},\ }\href {\doibase
  10.1103/PhysRevC.102.054911} {\bibfield  {journal} {\bibinfo  {journal}
  {Phys. Rev. C}\ }\textbf {\bibinfo {volume} {102}},\ \bibinfo {pages}
  {054911} (\bibinfo {year} {2020})},\ \Eprint
  {http://arxiv.org/abs/2004.02719} {arXiv:2004.02719 [nucl-th]} \BibitemShut
  {NoStop}%
\bibitem [{\citenamefont {Stephanov}(2009)}]{Stephanov:2008qz}%
  \BibitemOpen
  \bibfield  {author} {\bibinfo {author} {\bibfnamefont {M.~A.}\ \bibnamefont
  {Stephanov}},\ }\href {\doibase 10.1103/PhysRevLett.102.032301} {\bibfield
  {journal} {\bibinfo  {journal} {Phys. Rev. Lett.}\ }\textbf {\bibinfo
  {volume} {102}},\ \bibinfo {pages} {032301} (\bibinfo {year} {2009})},\
  \Eprint {http://arxiv.org/abs/0809.3450} {arXiv:0809.3450 [hep-ph]}
  \BibitemShut {NoStop}%
\bibitem [{\citenamefont {Asakawa}\ \emph {et~al.}(2009)\citenamefont
  {Asakawa}, \citenamefont {Ejiri},\ and\ \citenamefont
  {Kitazawa}}]{Asakawa:2009aj}%
  \BibitemOpen
  \bibfield  {author} {\bibinfo {author} {\bibfnamefont {M.}~\bibnamefont
  {Asakawa}}, \bibinfo {author} {\bibfnamefont {S.}~\bibnamefont {Ejiri}}, \
  and\ \bibinfo {author} {\bibfnamefont {M.}~\bibnamefont {Kitazawa}},\ }\href
  {\doibase 10.1103/PhysRevLett.103.262301} {\bibfield  {journal} {\bibinfo
  {journal} {Phys. Rev. Lett.}\ }\textbf {\bibinfo {volume} {103}},\ \bibinfo
  {pages} {262301} (\bibinfo {year} {2009})},\ \Eprint
  {http://arxiv.org/abs/0904.2089} {arXiv:0904.2089 [nucl-th]} \BibitemShut
  {NoStop}%
\bibitem [{\citenamefont {Stephanov}(2011)}]{Stephanov:2011pb}%
  \BibitemOpen
  \bibfield  {author} {\bibinfo {author} {\bibfnamefont {M.~A.}\ \bibnamefont
  {Stephanov}},\ }\href {\doibase 10.1103/PhysRevLett.107.052301} {\bibfield
  {journal} {\bibinfo  {journal} {Phys. Rev. Lett.}\ }\textbf {\bibinfo
  {volume} {107}},\ \bibinfo {pages} {052301} (\bibinfo {year} {2011})},\
  \Eprint {http://arxiv.org/abs/1104.1627} {arXiv:1104.1627 [hep-ph]}
  \BibitemShut {NoStop}%
\bibitem [{\citenamefont {Adam}\ \emph {et~al.}(2021)\citenamefont {Adam} \emph
  {et~al.}}]{STAR:2020tga}%
  \BibitemOpen
  \bibfield  {author} {\bibinfo {author} {\bibfnamefont {J.}~\bibnamefont
  {Adam}} \emph {et~al.} (\bibinfo {collaboration} {STAR}),\ }\href {\doibase
  10.1103/PhysRevLett.126.092301} {\bibfield  {journal} {\bibinfo  {journal}
  {Phys. Rev. Lett.}\ }\textbf {\bibinfo {volume} {126}},\ \bibinfo {pages}
  {092301} (\bibinfo {year} {2021})},\ \Eprint
  {http://arxiv.org/abs/2001.02852} {arXiv:2001.02852 [nucl-ex]} \BibitemShut
  {NoStop}%
\bibitem [{\citenamefont {Abdallah}\ \emph {et~al.}(2021)\citenamefont
  {Abdallah} \emph {et~al.}}]{STAR:2021iop}%
  \BibitemOpen
  \bibfield  {author} {\bibinfo {author} {\bibfnamefont {M.}~\bibnamefont
  {Abdallah}} \emph {et~al.} (\bibinfo {collaboration} {STAR}),\ }\href
  {\doibase 10.1103/PhysRevC.104.024902} {\bibfield  {journal} {\bibinfo
  {journal} {Phys. Rev. C}\ }\textbf {\bibinfo {volume} {104}},\ \bibinfo
  {pages} {024902} (\bibinfo {year} {2021})},\ \Eprint
  {http://arxiv.org/abs/2101.12413} {arXiv:2101.12413 [nucl-ex]} \BibitemShut
  {NoStop}%
\bibitem [{\citenamefont {Mukherjee}\ \emph {et~al.}(2016)\citenamefont
  {Mukherjee}, \citenamefont {Venugopalan},\ and\ \citenamefont
  {Yin}}]{Mukherjee:2016kyu}%
  \BibitemOpen
  \bibfield  {author} {\bibinfo {author} {\bibfnamefont {S.}~\bibnamefont
  {Mukherjee}}, \bibinfo {author} {\bibfnamefont {R.}~\bibnamefont
  {Venugopalan}}, \ and\ \bibinfo {author} {\bibfnamefont {Y.}~\bibnamefont
  {Yin}},\ }\href {\doibase 10.1103/PhysRevLett.117.222301} {\bibfield
  {journal} {\bibinfo  {journal} {Phys. Rev. Lett.}\ }\textbf {\bibinfo
  {volume} {117}},\ \bibinfo {pages} {222301} (\bibinfo {year} {2016})},\
  \Eprint {http://arxiv.org/abs/1605.09341} {arXiv:1605.09341 [hep-ph]}
  \BibitemShut {NoStop}%
\bibitem [{\citenamefont {An}\ \emph {et~al.}(2021)\citenamefont {An},
  \citenamefont {Basar}, \citenamefont {Stephanov},\ and\ \citenamefont
  {Yee}}]{An:2020vri}%
  \BibitemOpen
  \bibfield  {author} {\bibinfo {author} {\bibfnamefont {X.}~\bibnamefont
  {An}}, \bibinfo {author} {\bibfnamefont {G.}~\bibnamefont {Basar}}, \bibinfo
  {author} {\bibfnamefont {M.}~\bibnamefont {Stephanov}}, \ and\ \bibinfo
  {author} {\bibfnamefont {H.-U.}\ \bibnamefont {Yee}},\ }\href {\doibase
  10.1103/PhysRevLett.127.072301} {\bibfield  {journal} {\bibinfo  {journal}
  {Phys. Rev. Lett.}\ }\textbf {\bibinfo {volume} {127}},\ \bibinfo {pages}
  {072301} (\bibinfo {year} {2021})},\ \Eprint
  {http://arxiv.org/abs/2009.10742} {arXiv:2009.10742 [hep-th]} \BibitemShut
  {NoStop}%
\bibitem [{\citenamefont {Crossley}\ \emph
  {et~al.}(2017{\natexlab{a}})\citenamefont {Crossley}, \citenamefont
  {Glorioso},\ and\ \citenamefont {Liu}}]{Crossley:2015}%
  \BibitemOpen
  \bibfield  {author} {\bibinfo {author} {\bibfnamefont {M.}~\bibnamefont
  {Crossley}}, \bibinfo {author} {\bibfnamefont {P.}~\bibnamefont {Glorioso}},
  \ and\ \bibinfo {author} {\bibfnamefont {H.}~\bibnamefont {Liu}},\ }\href
  {\doibase 10.1007/JHEP09(2017)095} {\bibfield  {journal} {\bibinfo  {journal}
  {JHEP}\ }\textbf {\bibinfo {volume} {1709}},\ \bibinfo {pages} {095}
  (\bibinfo {year} {2017}{\natexlab{a}})},\ \Eprint
  {http://arxiv.org/abs/1511.03646} {arXiv:1511.03646 [hep-th]} \BibitemShut
  {NoStop}%
%%CITATION = arXiv:1511.03646;%%
\bibitem [{\citenamefont {Crossley}\ \emph
  {et~al.}(2017{\natexlab{b}})\citenamefont {Crossley}, \citenamefont
  {Glorioso},\ and\ \citenamefont {Liu}}]{Crossley:2017}%
  \BibitemOpen
  \bibfield  {author} {\bibinfo {author} {\bibfnamefont {M.}~\bibnamefont
  {Crossley}}, \bibinfo {author} {\bibfnamefont {P.}~\bibnamefont {Glorioso}},
  \ and\ \bibinfo {author} {\bibfnamefont {H.}~\bibnamefont {Liu}},\ }\href
  {\doibase 10.1007/JHEP09(2017)096} {\bibfield  {journal} {\bibinfo  {journal}
  {JHEP}\ }\textbf {\bibinfo {volume} {1709}},\ \bibinfo {pages} {096}
  (\bibinfo {year} {2017}{\natexlab{b}})},\ \Eprint
  {http://arxiv.org/abs/1701.07817} {arXiv:1701.07817 [hep-th]} \BibitemShut
  {NoStop}%
%%CITATION = arXiv:1701.07817;%%
\bibitem [{\citenamefont {Kovtun}\ \emph {et~al.}(2014)\citenamefont {Kovtun},
  \citenamefont {Moore},\ and\ \citenamefont {Romatschke}}]{Kovtun:2014hpa}%
  \BibitemOpen
  \bibfield  {author} {\bibinfo {author} {\bibfnamefont {P.}~\bibnamefont
  {Kovtun}}, \bibinfo {author} {\bibfnamefont {G.~D.}\ \bibnamefont {Moore}}, \
  and\ \bibinfo {author} {\bibfnamefont {P.}~\bibnamefont {Romatschke}},\
  }\href {\doibase 10.1007/JHEP07(2014)123} {\bibfield  {journal} {\bibinfo
  {journal} {JHEP}\ }\textbf {\bibinfo {volume} {07}},\ \bibinfo {pages} {123}
  (\bibinfo {year} {2014})},\ \Eprint {http://arxiv.org/abs/1405.3967}
  {arXiv:1405.3967 [hep-ph]} \BibitemShut {NoStop}%
%%CITATION = ARXIV:1405.3967;%%
\bibitem [{\citenamefont {Haehl}\ \emph {et~al.}(2016)\citenamefont {Haehl},
  \citenamefont {Loganayagam},\ and\ \citenamefont
  {Rangamani}}]{Haehl:2015foa}%
  \BibitemOpen
  \bibfield  {author} {\bibinfo {author} {\bibfnamefont {F.~M.}\ \bibnamefont
  {Haehl}}, \bibinfo {author} {\bibfnamefont {R.}~\bibnamefont {Loganayagam}},
  \ and\ \bibinfo {author} {\bibfnamefont {M.}~\bibnamefont {Rangamani}},\
  }\href {\doibase 10.1007/JHEP01(2016)184} {\bibfield  {journal} {\bibinfo
  {journal} {JHEP}\ }\textbf {\bibinfo {volume} {01}},\ \bibinfo {pages} {184}
  (\bibinfo {year} {2016})},\ \Eprint {http://arxiv.org/abs/1510.02494}
  {arXiv:1510.02494 [hep-th]} \BibitemShut {NoStop}%
\bibitem [{\citenamefont {Jensen}\ \emph {et~al.}(2018)\citenamefont {Jensen},
  \citenamefont {Pinzani-Fokeeva},\ and\ \citenamefont
  {Yarom}}]{Jensen:2017kzi}%
  \BibitemOpen
  \bibfield  {author} {\bibinfo {author} {\bibfnamefont {K.}~\bibnamefont
  {Jensen}}, \bibinfo {author} {\bibfnamefont {N.}~\bibnamefont
  {Pinzani-Fokeeva}}, \ and\ \bibinfo {author} {\bibfnamefont {A.}~\bibnamefont
  {Yarom}},\ }\href {\doibase 10.1007/JHEP09(2018)127} {\bibfield  {journal}
  {\bibinfo  {journal} {JHEP}\ }\textbf {\bibinfo {volume} {09}},\ \bibinfo
  {pages} {127} (\bibinfo {year} {2018})},\ \Eprint
  {http://arxiv.org/abs/1701.07436} {arXiv:1701.07436 [hep-th]} \BibitemShut
  {NoStop}%
\bibitem [{\citenamefont {Jain}\ and\ \citenamefont
  {Kovtun}(2020)}]{Jain:2020fsm}%
  \BibitemOpen
  \bibfield  {author} {\bibinfo {author} {\bibfnamefont {A.}~\bibnamefont
  {Jain}}\ and\ \bibinfo {author} {\bibfnamefont {P.}~\bibnamefont {Kovtun}},\
  }\href@noop {} {\  (\bibinfo {year} {2020})},\ \Eprint
  {http://arxiv.org/abs/2009.01356} {arXiv:2009.01356 [hep-th]} \BibitemShut
  {NoStop}%
\bibitem [{\citenamefont {Liu}\ and\ \citenamefont
  {Glorioso}(2018)}]{Glorioso:2018wxw}%
  \BibitemOpen
  \bibfield  {author} {\bibinfo {author} {\bibfnamefont {H.}~\bibnamefont
  {Liu}}\ and\ \bibinfo {author} {\bibfnamefont {P.}~\bibnamefont {Glorioso}},\
  }\bibfield  {booktitle} {\emph {\bibinfo {booktitle} {{Proceedings,
  Theoretical Advanced Study Institute in Elementary Particle Physics: Physics
  at the Fundamental Frontier (TASI 2017): Boulder, CO, USA, June 5-30,
  2017}}},\ }\href {\doibase 10.22323/1.305.0008} {\bibfield  {journal}
  {\bibinfo  {journal} {PoS}\ }\textbf {\bibinfo {volume} {TASI2017}},\
  \bibinfo {pages} {008} (\bibinfo {year} {2018})},\ \Eprint
  {http://arxiv.org/abs/1805.09331} {arXiv:1805.09331 [hep-th]} \BibitemShut
  {NoStop}%
%%CITATION = ARXIV:1805.09331;%%
\bibitem [{\citenamefont {Dubovsky}\ \emph {et~al.}(2012)\citenamefont
  {Dubovsky}, \citenamefont {Hui}, \citenamefont {Nicolis},\ and\ \citenamefont
  {Son}}]{Dubovsky:2011sj}%
  \BibitemOpen
  \bibfield  {author} {\bibinfo {author} {\bibfnamefont {S.}~\bibnamefont
  {Dubovsky}}, \bibinfo {author} {\bibfnamefont {L.}~\bibnamefont {Hui}},
  \bibinfo {author} {\bibfnamefont {A.}~\bibnamefont {Nicolis}}, \ and\
  \bibinfo {author} {\bibfnamefont {D.~T.}\ \bibnamefont {Son}},\ }\href
  {\doibase 10.1103/PhysRevD.85.085029} {\bibfield  {journal} {\bibinfo
  {journal} {Phys. Rev. D}\ }\textbf {\bibinfo {volume} {85}},\ \bibinfo
  {pages} {085029} (\bibinfo {year} {2012})},\ \Eprint
  {http://arxiv.org/abs/1107.0731} {arXiv:1107.0731 [hep-th]} \BibitemShut
  {NoStop}%
\bibitem [{\citenamefont {Glorioso}\ \emph {et~al.}(2017)\citenamefont
  {Glorioso}, \citenamefont {Crossley},\ and\ \citenamefont
  {Liu}}]{Glorioso:2017fpd}%
  \BibitemOpen
  \bibfield  {author} {\bibinfo {author} {\bibfnamefont {P.}~\bibnamefont
  {Glorioso}}, \bibinfo {author} {\bibfnamefont {M.}~\bibnamefont {Crossley}},
  \ and\ \bibinfo {author} {\bibfnamefont {H.}~\bibnamefont {Liu}},\ }\href
  {\doibase 10.1007/JHEP09(2017)096} {\bibfield  {journal} {\bibinfo  {journal}
  {JHEP}\ }\textbf {\bibinfo {volume} {09}},\ \bibinfo {pages} {096} (\bibinfo
  {year} {2017})},\ \Eprint {http://arxiv.org/abs/1701.07817} {arXiv:1701.07817
  [hep-th]} \BibitemShut {NoStop}%
\bibitem [{\citenamefont {Glorioso}\ and\ \citenamefont
  {Liu}(2016)}]{Glorioso:2016gsa}%
  \BibitemOpen
  \bibfield  {author} {\bibinfo {author} {\bibfnamefont {P.}~\bibnamefont
  {Glorioso}}\ and\ \bibinfo {author} {\bibfnamefont {H.}~\bibnamefont {Liu}},\
  }\href@noop {} {\  (\bibinfo {year} {2016})},\ \Eprint
  {http://arxiv.org/abs/1612.07705} {arXiv:1612.07705 [hep-th]} \BibitemShut
  {NoStop}%
\bibitem [{\citenamefont {Sogabe}\ \emph {et~al.}(2021)\citenamefont {Sogabe},
  \citenamefont {Yamamoto},\ and\ \citenamefont {Yin}}]{sogabe2021positive}%
  \BibitemOpen
  \bibfield  {author} {\bibinfo {author} {\bibfnamefont {N.}~\bibnamefont
  {Sogabe}}, \bibinfo {author} {\bibfnamefont {N.}~\bibnamefont {Yamamoto}}, \
  and\ \bibinfo {author} {\bibfnamefont {Y.}~\bibnamefont {Yin}},\ }\href@noop
  {} {\enquote {\bibinfo {title} {Positive magnetoresistance induced by
  hydrodynamic fluctuations in chiral media},}\ } (\bibinfo {year} {2021}),\
  \Eprint {http://arxiv.org/abs/2105.10271} {arXiv:2105.10271 [hep-th]}
  \BibitemShut {NoStop}%
\bibitem [{\citenamefont {Wang}\ and\ \citenamefont
  {Heinz}(2002)}]{Wang:1998wg}%
  \BibitemOpen
  \bibfield  {author} {\bibinfo {author} {\bibfnamefont {E.}~\bibnamefont
  {Wang}}\ and\ \bibinfo {author} {\bibfnamefont {U.~W.}\ \bibnamefont
  {Heinz}},\ }\href {\doibase 10.1103/PhysRevD.66.025008} {\bibfield  {journal}
  {\bibinfo  {journal} {Phys. Rev. D}\ }\textbf {\bibinfo {volume} {66}},\
  \bibinfo {pages} {025008} (\bibinfo {year} {2002})},\ \Eprint
  {http://arxiv.org/abs/hep-th/9809016} {arXiv:hep-th/9809016} \BibitemShut
  {NoStop}%
\bibitem [{\citenamefont {Janssen}\ \emph {et~al.}(1989)\citenamefont
  {Janssen}, \citenamefont {Schaub},\ and\ \citenamefont
  {Schmittmann}}]{initial-slip}%
  \BibitemOpen
  \bibfield  {author} {\bibinfo {author} {\bibfnamefont {H.~K.}\ \bibnamefont
  {Janssen}}, \bibinfo {author} {\bibfnamefont {B.}~\bibnamefont {Schaub}}, \
  and\ \bibinfo {author} {\bibfnamefont {B.}~\bibnamefont {Schmittmann}},\
  }\href {\doibase 10.1007/BF01319383} {\bibfield  {journal} {\bibinfo
  {journal} {Zeitschrift f{\"u}r Physik B Condensed Matter}\ }\textbf {\bibinfo
  {volume} {73}},\ \bibinfo {pages} {539} (\bibinfo {year} {1989})}\BibitemShut
  {NoStop}%
\bibitem [{\citenamefont {Pradeep}\ \emph {et~al.}(2021)\citenamefont
  {Pradeep}, \citenamefont {Rajagopal}, \citenamefont {Stephanov},\ and\
  \citenamefont {Yin}}]{Pradeep:2021opj}%
  \BibitemOpen
  \bibfield  {author} {\bibinfo {author} {\bibfnamefont {M.}~\bibnamefont
  {Pradeep}}, \bibinfo {author} {\bibfnamefont {K.}~\bibnamefont {Rajagopal}},
  \bibinfo {author} {\bibfnamefont {M.}~\bibnamefont {Stephanov}}, \ and\
  \bibinfo {author} {\bibfnamefont {Y.}~\bibnamefont {Yin}},\ }in\ \href@noop
  {} {\emph {\bibinfo {booktitle} {{International Conference on Critical Point
  and Onset of Deconfinement}}}}\ (\bibinfo {year} {2021})\ \Eprint
  {http://arxiv.org/abs/2109.13188} {arXiv:2109.13188 [hep-ph]} \BibitemShut
  {NoStop}%
\bibitem [{\citenamefont {Oliinychenko}\ and\ \citenamefont
  {Koch}(2019)}]{Oliinychenko:2019zfk}%
  \BibitemOpen
  \bibfield  {author} {\bibinfo {author} {\bibfnamefont {D.}~\bibnamefont
  {Oliinychenko}}\ and\ \bibinfo {author} {\bibfnamefont {V.}~\bibnamefont
  {Koch}},\ }\href {\doibase 10.1103/PhysRevLett.123.182302} {\bibfield
  {journal} {\bibinfo  {journal} {Phys. Rev. Lett.}\ }\textbf {\bibinfo
  {volume} {123}},\ \bibinfo {pages} {182302} (\bibinfo {year} {2019})},\
  \Eprint {http://arxiv.org/abs/1902.09775} {arXiv:1902.09775 [hep-ph]}
  \BibitemShut {NoStop}%
\bibitem [{\citenamefont {Oliinychenko}\ \emph {et~al.}(2020)\citenamefont
  {Oliinychenko}, \citenamefont {Shi},\ and\ \citenamefont
  {Koch}}]{Oliinychenko:2020cmr}%
  \BibitemOpen
  \bibfield  {author} {\bibinfo {author} {\bibfnamefont {D.}~\bibnamefont
  {Oliinychenko}}, \bibinfo {author} {\bibfnamefont {S.}~\bibnamefont {Shi}}, \
  and\ \bibinfo {author} {\bibfnamefont {V.}~\bibnamefont {Koch}},\ }\href
  {\doibase 10.1103/PhysRevC.102.034904} {\bibfield  {journal} {\bibinfo
  {journal} {Phys. Rev. C}\ }\textbf {\bibinfo {volume} {102}},\ \bibinfo
  {pages} {034904} (\bibinfo {year} {2020})},\ \Eprint
  {http://arxiv.org/abs/2001.08176} {arXiv:2001.08176 [hep-ph]} \BibitemShut
  {NoStop}%
\bibitem [{\citenamefont {Gao}\ \emph {et~al.}(2020)\citenamefont {Gao},
  \citenamefont {Glorioso},\ and\ \citenamefont {Liu}}]{Gao:2018bxz}%
  \BibitemOpen
  \bibfield  {author} {\bibinfo {author} {\bibfnamefont {P.}~\bibnamefont
  {Gao}}, \bibinfo {author} {\bibfnamefont {P.}~\bibnamefont {Glorioso}}, \
  and\ \bibinfo {author} {\bibfnamefont {H.}~\bibnamefont {Liu}},\ }\href
  {\doibase 10.1007/JHEP03(2020)040} {\bibfield  {journal} {\bibinfo  {journal}
  {JHEP}\ }\textbf {\bibinfo {volume} {03}},\ \bibinfo {pages} {040} (\bibinfo
  {year} {2020})},\ \Eprint {http://arxiv.org/abs/1803.10778} {arXiv:1803.10778
  [hep-th]} \BibitemShut {NoStop}%
\end{thebibliography}%

\end{document}